\documentclass[a4paper,11pt]{article}
\pdfoutput=1 

\usepackage{jheppub} 

\usepackage[T1]{fontenc} 
\usepackage{amssymb, bm, graphicx, graphics, color, mathrsfs, multirow}
\usepackage{subfigure} \subfiguretopcaptrue

\newcommand{\be}{\begin{equation}}
\newcommand{\ee}{\end{equation}}
\newcommand{\ben}{\begin{eqnarray}}
\newcommand{\een}{\end{eqnarray}}

\newcommand{\la}{{\lambda}}

\newcommand{\cL}{{\cal L}}

\newcommand{\na}{\nabla}

\newcommand{\tpe}{{\tilde p}}

\newcommand{\hg}{\hat g}
\newcommand{\hR}{\hat R}
\newcommand{\hD}{\hat D}

\newcommand{\hna}{\hat \nabla}

\newcommand{\zpsi}{\psi^{\ast}}

\newcommand{\hS}{\hat S}

\newcommand{\as}{\tpe_s}
\newcommand{\ah}{\tpe_h}

\newcommand{\POv}{_{,v}}
\newcommand{\POu}{_{,u}}
\newcommand{\POuv}{_{,uv}}

\makeatletter
\newcommand{\raisemath}[1]{\mathpalette{\raisem@th{#1}}}
\newcommand{\raisem@th}[3]{\raisebox{#1}{$#2#3$}}
\makeatother

\title{\boldmath Gravitational collapse involving electric charge in the decoupling limit of the dilatonic Gauss--Bonnet gravity}

\author{Anna Nakonieczna}
\author{and {\L}ukasz Nakonieczny}

\affiliation{Institute of Theoretical Physics, Faculty of Physics, University of Warsaw, \\
Pasteura 5, 02-093 Warszawa, Poland}

\emailAdd{Anna.Nakonieczna@fuw.edu.pl}
\emailAdd{Lukasz.Nakonieczny@fuw.edu.pl}

\abstract{
The paper presents the course and outcomes of gravitational collapse of a~self-interacting electrically charged scalar field in the decoupling limit of the dilatonic Gauss--Bonnet gravity. The~analyses were conducted with the use of double null coordinates, which allow to trace the evolutions within the whole spacetime, i.e., from approximately past null infinity up to the central hypersurface, which can be singular and surrounded by a~horizon. The~emerging spacetimes were either non-singular or, for sufficiently big scalar fields self-interaction strengths, singular and containing black holes of a~Schwarzschild type. A~size of the region, in which the gravitational dynamics was observed was controlled by an absolute value of the Gauss--Bonnet coupling. Dependencies of characteristics of the forming black holes on the dilatonic and Gauss--Bonnet parameters turned out to be similar in the case of black hole masses and radii as well as their time of formation in terms of retarded time. In~the cases of masses and radii minima were observed, while in the remaining case a~maximum existed. The~electric charge of the emerging black holes possessed a~maximum when measured versus the dilatonic coupling constant and was strictly decreasing with the Gauss--Bonnet coupling. All the characteristics changed monotonically with the field self-interaction strengths. The~times of formation and charges of black holes decreased, while masses and radii increased with the self-interaction strengths of the dynamical fields. Values of the energy density, radial pressure, pressure anisotropy and the collapsing scalar fields were the biggest along the null hypersurface of propagation of the initial peaks of the scalar fields. For big values of the Gauss--Bonnet coupling constant, an increase in their values was also observed in the vicinity of the central singularity within the whole range of advanced time. Non-zero values of the dilaton field outside the black hole event horizon may indicate a~formation of a~hairy black hole. The~local temperature calculated along the apparent horizon was increasing for late times of the evolution, that is in non-dynamical spacetime regions, and exhibited extrema in areas, where the dynamics of the gravity-matter system was observed.
}

\begin{document} 
\maketitle
\flushbottom

\section{Introduction}
\label{sec:intro}

Dynamical gravitational collapse is a~fully non-linearly and non-perturbatively described process in which a~mutual evolution of the spacetime geometry and matter content of the evolving spacetime is tracked. It~means that these two features affect each other in the course of the process and hence the described outcomes are closer to reality, but at the same time more difficult to obtain than in the case of matter evolutions on predefined backgrounds, even if backreaction is considered. When an electrically charged scalar field is included in the examined matter-geometry system, the whole system can be regarded as simulating real astrophysical evolutions. Thus obtained spacetime structures satisfactorily resemble the structures anticipated in the case of a~non-charged, but rotating collapse, which takes place during real astrophysical events~\cite{SorkinPiran2001-084006,SorkinPiran2001-124024}.

Up to now, structures of spacetimes emerging from the dynamical gravitational collapse were extensively studied first in the simplest cases of sole self-interacting neutral~\cite{HamadeStewart1996-497} and electrically charged~\cite{HodPiran1998-1554,HodPiran1998-1555,OrenPiran2003-044013,HongHwangStewartYeom2010-045014,HwangYeom2011-064020} scalar fields. More sophisticated theories, complicated in either or both gravitational and matter sectors were studied in~\cite{HansenHwangYeom2009-016,HwangYeom2010-205002,HwangYeom2011-155003,HwangLeeYeom2011-006,BorkowskaRogatkoModerski2011-084007,NakoniecznaRogatko2012-3175,NakoniecznaRogatkoModerski2012-044043,NakoniecznaRogatko-AIP,HansenYeom2014-040,ChenYeom2015-022,HansenYeom2015-019,NakoniecznaRogatkoNakonieczny2015-012}. Apart from investigating structures of spacetimes forming in the process of interest, additional problems related to gravitational dynamics, such as pair creation during collapse and subsequent evaporation of black holes, issues of the cosmic censorship conjecture and the information loss problem, the impact of changing the number of dimensions and the role of various topologies, time measurements as well as observables, were discussed in~\cite{SorkinPiran2001-084006,SorkinPiran2001-124024,HansenHwangYeom2009-016,HongHwangStewartYeom2010-045014,HwangYeom2011-064020,HwangYeom2011-155003,HwangKimYeom2012-055003,HwangLeeLeeYeom2012-003,HwangLeeParkYeom2012-003,NakoniecznaLewandowski2015-064031,NakoniecznaYeom2016-049,NakoniecznaYeom2016-155,NakoniecznaNakonieczny2020-1051}. The~up-to-date summary of the sketched researches can be found in~\cite{NakoniecznaNakoniecznyYeom2019-1930006}.

The Einstein--dilaton--Gauss--Bonnet (EdGB) theory is a~beyond general relativity formulation, which is a~scalar--tensor theory including higher order terms in curvature. It~has been extensively studied in various cosmological large-scale contexts such as dark energy explanations, inflation or early universe bouncing models~\cite{Quiros2019-1930012,Kobayashi2019-086901}. It~has been also involved in small-scale astrophysical analyses like investigating black holes and their mergers or creation and stability of wormholes~\cite{CuyubambaKonoplyaZhidenko2018-044040,TumurtushaaYeom2019-488,WitekGualtieriPaniSotiriou2019-064035,TumurtushaaYeom2020-arxiv}. Experimental predictions of the EdGB theory are consistent with current general relativity tests, but the theory yields different results for mergers of the stellar mass black holes~\cite{YagiSteinYunes2016-024010}, what makes it interesting in the context of analyses of the recently extremely viable gravitational waves observations.

So far, a~spherical gravitational collapse in the EdGB theory was considered within its shift symmetric version in~\cite{RipleyPretorius2019-134001}. The~research was conducted in coordinates which do not penetrate the horizon, hence the collapse results were described up to the forming horizon and its interior was not analysed. This successful attempt to describe the outcomes of a~fully non-linear process within the EdGB theory was an introductory step towards analysis of the whole spacetime, including interiors of emerging singular objects. Our current studies present another approach to this problem, namely the engaged coordinates cover the whole spacetime, yet their usage also requires introducing a~simplification of the full EdGB theory.

The decoupling limit of the EdGB theory is a~version of the full theory, in which the Gauss--Bonnet term is not included in the derivation of the Einstein equations, only in the equations describing the matter sector of the theory. This means that the scalar sector of the theory does not backreact on geometry~\cite{WitekGualtieriPaniSotiriou2019-064035}. This simplified formulation of the full EdGB theory has been recently employed in research on black hole hair formation and scalar modes~\cite{BenkelSotiriouWitek2016-121503R,BenkelSotiriouWitek2017-064001,BlazquezSalcedoKhooKunz2017-064008}. Investigating the collapse within the truncated version of the EdGB theory will be performed in double null coordinates, which enable to trace the gravitational evolution within the whole spacetime, that is from approximately null infinity up to the central singularity. This is the first step towards investigations of the course and results of gravitational evolutions in the full version of the EdGB theory in the whole forming spacetimes, not only regions exterior to the nascent singular objects.

The paper is organized as follows. In~section~\ref{sec:model} the considered model was presented. Section~\ref{sec:particulars} contains details of numerical calculations and results analysis. The~obtained results were described and discussed in sections~\ref{sec:struc}--\ref{sec:obs}. Section~\ref{sec:conclusions} summarizes the undertaken research. Technical details of the numerical code preparation and its tests are presented in appendix~\ref{sec:appendix}.

\section{Theoretical model of the evolution}
\label{sec:model}

The studied theoretical model of the dynamical collapse involves an electrically charged scalar field collapsing within the decoupling limit of the dilaton--Gauss--Bonnet gravity. The~general form of the action written, due to the fact that the whole construction involves the string theory concepts, in the string frame is the following:
\ben
\hS = \int d^{4} x \sqrt{-\hg} \left\{ e^{- 2 \phi}
\left[ \hR - 2 \left( \hna \phi \right)^{\raisemath{-1.15pt}{\hspace{-0.05cm}2}} + e^{2 \alpha \phi} \left( \hat{\cL}_{SF}  
+ \gamma \hat{\cL}_{GB} \right) \right] \right\},
\label{a1}
\een
where $\phi$ stands for the dilaton field, $\alpha$ is the dilatonic coupling constant and $\gamma$ determines the coupling between the Gauss--Bonnet contribution and the dilaton. The~geometrized units system, in which \mbox{$8\pi G=c=1$}, was used in the computations. The~Lagrangian of the electrically charged scalar field $\psi$ is given by the expression
\ben
\hat{\cL}_{SF} = - \frac{1}{2} \hD_{\beta} \psi \left(\hD^\beta \psi\right)^{\raisemath{-1pt}{\hspace{-0.05cm}\ast}} 
- F_{\beta \sigma} F^{\beta \sigma},
\een
where $F_{\beta \sigma}$ is the Maxwell field strength tensor. The~covariant derivative is $\hD_\beta = \hna_{\beta} + ieA_{\beta}$, where $e$ is the electric coupling constant, $A_\beta$ is the four-potential and $i$ denotes the imaginary unit. The~Gauss--Bonnet Lagrangian is defined in a~standard way
\ben
\hat{\cL}_{GB} = \hat{R}^2 - 4\hat{R}_{\beta\sigma}\hat{R}^{\beta\sigma} + \hat{R}_{\beta\sigma\gamma\delta}\hat{R}^{\beta\sigma\gamma\delta}.
\een

The variation of the constructed action~\eqref{a1} with respect to matter fields, namely the dilaton field $\phi$, the Maxwell field $A_\mu$ and the complex scalar field $\psi$ results in the following set of evolution equations:
\ben
\na^{2} \phi -\frac{\alpha+1}{4} e^{ 2\phi \left(\alpha+1\right)} D_\nu \psi \left( D^\nu \psi \right)^\ast
- \frac{1}{2}\alpha e^{ 2\alpha \phi} F_{\beta \sigma} F^{\beta \sigma} +2\gamma\alpha e^{2\alpha\phi}\hat{\cL}_{GB} &= 0,
\label{aaa} \\
 \na_{\mu} \left( e^{ 2\alpha \phi} F^{\mu \nu} \right) 
+ \frac{1}{4} e^{ 2\phi \left(\alpha+1\right)} \Big[
i e \zpsi D^\nu \psi
- i e \psi \left( D^\nu \psi \right)^\ast \Big] &= 0,
\label{bbb} \\
\na^{2} \psi + i e A^{\beta} \left(
2 \na_{\beta} + i e A_{\beta} \right) \psi
+ i e \na_{\beta} A^{\beta} \psi &= 0,
\label{ccc} \\
\na^{2} \zpsi - i e A^{\beta} \left(
2 \na_{\beta} - i e A_{\beta} \right) \zpsi
- i e \na_{\beta} A^{\beta} \zpsi &= 0.
\label{c1c1c1}
\een
The derivation of the above equations of motion involved a~conversion of the string frame into the Einstein frame, which are related via the conformal transformation
\ben
g_{\mu \nu} = e^{- 2 \phi} \hg_{\mu \nu},
\label{eqn:conf-tr}
\een
where $g_{\mu \nu}$ and $\hg_{\mu \nu}$ denote metric tensors in the Einstein and string frames, respectively~\cite{Ortin}. The~transformation between the two frames preserves causality. Since the obtained results will be mainly elaborated on employing notions related to causal structures of emerging spacetimes and features of the intrinsic dynamical objects, we do not expect the transformation to influence the ultimate conclusions and hence they can be regarded as physically relevant. From now on, all variables and quantities are written in the Einstein frame.

The gravitational field equations derived by varying the action~\eqref{a1} with respect to the metric tensor, taking the truncation of the theory into account, will complement the above set of equations describing the examined dynamical system. For the current studies, the double null spherically symmetric line element~\cite{MisnerThorneWheeler} is chosen
\ben
ds^2 = - a(u, v)^2 du dv + r^2(u, v) d \Omega^2,
\label{m}
\een
where $u$ and $v$ are retarded and advanced time null coordinates, respectively, and $d \Omega^2 = d\Theta^2 + \sin^2\Theta d\Phi^2 $ is the line element of the unit sphere, where $\Theta$ and $\Phi$ are angular coordinates. This choice of coordinates determines the spacetime foliation for conducting computations, which becomes 2+2~\cite{InvernoSmallwood1980-1223}. The~double null formalism has been widely and successfully employed in dynamical gravitational collapse researches, e.g.,~\cite{SorkinPiran2001-084006,SorkinPiran2001-124024,HodPiran1998-1554,OrenPiran2003-044013,HongHwangStewartYeom2010-045014,HwangYeom2011-064020}. Its~advantage is that it enables to follow the dynamical evolution from approximately past null infinity, through the formation of possible horizons up to the final central singularity of singular spacetimes.

In spherical symmetry the only non-vanishing components of the electromagnetic field tensor are $F_{uv}$ and $F_{vu}$. Due to the gauge freedom $A_{u} \to A_{u} + \na_{u} \theta^\prime$, where $\theta^\prime = \int A_{v}dv$, the only non-zero four-vector component is $A_{u}$. It~is a~function of retarded and advanced time.

The dilaton field equation of motion~\eqref{aaa} in the chosen coordinate system is given by
\ben
r\POu \phi\POv + r\POv \phi\POu + r \phi\POuv -\alpha e^{2\alpha\phi} {Q^2 a^2 \over 4 r^3} +&\nonumber\\
-\frac{\alpha+1}{8}r e^{ 2\phi \left(\alpha+1\right)}
\Big[\psi\POu\zpsi\POv+\psi\POv\zpsi\POu +ieA_u\left(\psi\zpsi\POv-\zpsi\psi\POv\right)\Big] +&\nonumber\\
-16\gamma\alpha e^{2\alpha\phi} \frac{1}{a^4 r} \Big[12a\POu a\POv r\POu r\POv - 4a\left(a\POv r\POv r_{,uu} + a\POu r\POu r_{,vv}\right) +&\nonumber\\
+2a^2 \left(r_{,uu}r_{,vv}-r_{,uv}^2\right) -4aa_{,uv}r\POu r\POv +a^2\left(a\POu a\POv - aa_{,uv}\right) \Big] &= 0,
\label{d}
\een
where we set
\be
Q = 2 {A_{u, v} r^2 \over a^2}.
\label{charge}
\ee
$Q$ is a~function of retarded and advanced time and determines the electric charge within a~sphere of a~radius $r(u,v)$ on a~spacelike hypersurface
containing the point $(u,v)$. Partial derivatives with respect to the null coordinates are marked as $\POu$ and $\POv$. Concerning the assumed line element~\eqref{m} and the definition of electric charge~\eqref{charge}, the $v$-component of Maxwell equations~\eqref{bbb} can be separated into two first-order differential equations. The~first one governs the evolution of the only non-zero component of the four-vector of the Maxwell field
\ben
A_{u, v} - {Q a^{2} \over 2 r^{2}} = 0,
\label{p}
\een
and the second one describes the behavior of the quantity $Q$ during the evolution
\ben
Q\POv + 2\alpha \phi\POv Q + {i e r^2 \over 4} e^{2 \phi}
\left( \zpsi \psi\POv - \psi \zpsi\POv \right) = 0,
\label{l}
\een
The equations for the complex scalar field~\eqref{ccc}--\eqref{c1c1c1} become
\ben
r\POu \psi\POv + r\POv \psi\POu + r \psi\POuv
+ i e r A_{u}\psi\POv + ier\POv A_{u}\psi 
+ {ieQa^2 \over 4 r} \psi &= 0,
\label{s1} \\
r\POu \zpsi\POv + r\POv \zpsi\POu + r \zpsi\POuv 
- i e r A_{u} \zpsi\POv - i e r\POv A_{u} \zpsi 
- {i e Q a^2 \over 4 r} \zpsi &= 0.
\label{s2}
\een

The stress-energy tensor for the considered electrically charged scalar field evolving within the decoupling limit of the dilaton-Gauss--Bonnet theory is the following:
\ben
T_{\mu \nu} &=& 2\phi_{,\mu} \phi_{,\nu} - g_{\mu \nu} \phi_{,\beta}\phi^{,\beta} + e^{2\alpha\phi} \bigg(
2 F_{\mu \beta} F_{\nu}{}{}^{\beta} - {1 \over 2} g_{\mu \nu} F_{\beta \sigma} F^{\beta \sigma} \bigg) + \nonumber \\
&& - \frac{1}{4} e^{ 2\phi \left(\alpha+1\right)} \bigg[ g_{\mu \nu} D_\beta \psi \left(D^\beta \psi\right)^{\raisemath{-1pt}{\hspace{-0.05cm}\ast}}
- D_\mu \psi \left(D_\nu \psi\right)^\ast - \left(D_\mu \psi\right)^\ast D_\nu \psi \bigg].
\label{ten}
\een
Its non-vanishing components written in double null coordinates are
\ben
\label{eqn:Tuu}
T_{uu} &=& 2\phi\POu^2 + \frac{1}{2} e^{ 2\phi \left(\alpha+1\right)}
\Big[\psi\POu\zpsi\POu+ieA_u\left(\psi\zpsi\POu-\zpsi\psi\POu\right)+e^2A_u^2\psi\zpsi\Big], \\
T_{vv} &=& 2\phi\POv^2 + \frac{1}{2} e^{ 2\phi \left(\alpha+1\right)} \psi\POv\zpsi\POv, \\
T_{uv} &=& e^{2\alpha \phi} \frac{Q^2a^2}{2r^4}, \\
T_{\theta\theta} &=& 4\frac{r^2}{a^2}\phi\POu\phi\POv + e^{2\alpha \phi} \frac{Q^2}{r^2} + \nonumber\\
&&+\frac{1}{2}\frac{r^2}{a^2} e^{ 2\phi \left(\alpha+1\right)} \Big[\psi\POu\zpsi\POv+\psi\POv\zpsi\POu +ieA_u\left(\psi\zpsi\POv-\zpsi\psi\POv\right)\Big].
\label{eqn:Ttt}
\een
Combining the adequate components of the Einstein tensor resulting from the metric~\eqref{m} and the obtained stress-energy tensor components, the Einstein equations yield
{\allowdisplaybreaks
\ben
{2 a\POu r\POu \over a} - r_{,u u}
&=& r \phi\POu^2 + {r \over 4} e^{ 2\phi \left(\alpha+1\right)} \Big[
\psi\POu \zpsi\POu + i e A_{u} \left(
\psi\zpsi\POu - \zpsi\psi\POu \right) + e^2 A_{u}^2 \psi \zpsi \Big], \hspace{1cm}
\label{e1} \\
{2 a\POv r\POv \over a} - r_{,vv} &=&
r \phi\POv^2 + {r \over 4} e^{ 2\phi \left(\alpha+1\right)} \psi\POv \zpsi\POv, \\
{a^2 \over 4r} + {r\POu r_{v} \over r} + r\POuv &=& e^{2\alpha \phi}{Q^2 a^2 \over 4 r^3}, \\
{a\POu a\POv \over a^2}
- {a\POuv \over a} - {r\POuv \over r} &=&
e^{2\alpha \phi} {Q^2 a^2 \over 4 r^4} + \phi\POu \phi\POv +\nonumber\\
&&+ {1 \over 8} e^{ 2\phi \left(\alpha+1\right)} \Big[
\psi\POu \zpsi\POv + \zpsi\POu \psi\POv
+ i e A_{u} \left(\psi \zpsi_{v} - \zpsi\psi\POv \right) \Big].
\label{e4}
\een}
They complement the matter equations presented above in order to obtain the complete set of equations describing the dynamics of the examined system, which are~\eqref{d}--\eqref{s2}, without the relation~\eqref{charge}, which is a~definition of the physical quantity $Q$.

The following set of auxiliary variables is introduced to prepare the obtained set of dynamical equations to be solved numerically:
\ben\label{eqn:substitution}
\begin{split}
c &= \frac{a\POu}{a}, \qquad &d&= \frac{a\POv}{a},
\qquad &f&= r\POu, \qquad &g&= r\POv, \\
h &= \phi, \qquad &x&= \phi\POu, \qquad &y&= \phi\POv, \\
s &= \psi, \qquad &p&= \psi\POu, \qquad &q&= \psi\POv, \qquad &\beta&= A_u.
\end{split}
\een
It is supplemented by the quantities
\ben
\la \equiv \frac{a^2}{4} + f g, \qquad \mu \equiv f q + g p, \qquad \eta \equiv gx + fy.
\een
The above variables make it possible to rewrite the second-order differential equations~\eqref{d}, \eqref{s1}--\eqref{s2} and~\eqref{e1}--\eqref{e4} as first-order ones. Additionally, real fields $\psi_1$ and $\psi_2$ are introduced instead of conjugate fields $\psi$ and $\zpsi$ according to relations $\psi = \psi_{1} + i \psi_{2}$, $\zpsi = \psi_{1} - i \psi_{2}$, what results in
\ben
\begin{split}
s &= s_1 + i s_2, \qquad &p& = p_1 + i p_2, \qquad &q& = q_1 + i q_2, \\
\mu &= \mu_1 + i \mu_2, \qquad &\mu_1& = f q_1 + g p_1, \qquad &\mu_2& = fq_2 + gp_2.
\label{defdef}
\end{split}
\een

After introducing the above substitutions, the final system of equations of motion describing the gravitational collapse of interest can be written as
{\allowdisplaybreaks
\ben \label{eqn:P1-2}
P1: && a\POu = ac,\\
P2: && a\POv = ad,\\
P3: && r\POu = f,\\
P4: && r\POv = g,\\
P5: && s_{1(2),u} = p_{1(2)},\\
P6: && s_{1(2),v} = q_{1(2)},\\
P7: && h\POu = x,\\
P8: && h\POv = y,\\
E1: && f\POu = 2 c f - r x^2 - \frac{r}{4} e^{2h\left(\alpha+1\right)}
\Big[p_1^{\: 2} + p_2^{\: 2} + 2e\beta \left(s_1 p_2 - s_2 p_1 \right)
+ e^2 \beta^2 \left(s_1^{\: 2} + s_2^{\: 2} \right) \Big],
\label{eqn:E1} \\
E2: && g\POv = 2 d g - r y^2 - \frac{r}{4} e^{2h\left(\alpha+1\right)} \left(q_1^{\: 2} + q_2^{\: 2}\right), \\
E3: && g\POu = f\POv = - \frac{\la}{r} + e^{2\alpha h} \frac{Q^2 a^2}{4 r^3}, \\
E4: && d\POu = c\POv = \frac{\la}{r^2} - xy -\frac{1}{4} e^{2h \left(\alpha + 1\right)}
\Big[p_1 q_1 + p_2 q_2 + e\beta \left(s_1 q_2 - s_2 q_1 \right) \Big] - e^{2\alpha h} \frac{Q^2 a^2}{2 r^4}, \hspace{0.95cm}
\label{eqn:E4} \\
S_{_{\left(Re\right)}}: && q_{1,u} = p_{1,v} = - \frac{\mu_1}{r} 
+ e \beta q_2 + e s_2 \beta \frac{g}{r} + e s_2 \frac{Qa^2}{4r^2}, \\
S_{_{\left(Im\right)}}: && q_{2,u} = p_{2,v} = - \frac{\mu_2}{r} 
- e \beta q_1 - e s_1 \beta \frac{g}{r} - e s_1 \frac{Qa^2}{4r^2}, \\
M1: && \beta\POv = \frac{Qa^2}{2 r^2},\\
M2: && Q\POv = -2\alpha y Q + \frac{e r^2}{2} e^{2h} \left( s_1 q_2 - s_2 q_1 \right),
\label{eqn:M2} \\
D: && y\POu = x\POv = -\frac{\eta}{r} + \alpha e^{2\alpha h}\frac{Q^2a^2}{4r^4} \left[ 1+ \frac{32}{a^2r}\gamma e^{2\alpha h} \left( 6\lambda - e^{2\alpha h} \frac{Q^2a^2}{4r^2} \right) \right] + \nonumber\\
&& + \left( \frac{\alpha+1}{4} +\frac{16}{a^2r}\lambda\gamma\alpha e^{2\alpha h} \right) e^{2h\left(\alpha+1\right)}
\Big[p_1q_1 + p_2q_2 + e\beta\left(s_1q_2 - s_2q_1 \right) \Big] + \nonumber \\
&& + \frac{32}{a^2r}\gamma\alpha e^{2\alpha h} \left\{ \lambda \left(2xy - 3\frac{\lambda}{r^2}\right) + r^2\left[ x^2y^2 + \frac{x^2}{4} e^{2h\left(\alpha+1\right)} \left(q_1^{\: 2} + q_2^{\: 2}\right) \right.\right. + \nonumber\\
&& \left.\left. + \frac{1}{4} e^{2h\left(\alpha+1\right)} \left(y^2 + \frac{1}{4} e^{2h\left(\alpha+1\right)} \left(q_1^{\: 2} + q_2^{\: 2}\right)\right) \cdot \right.\right. \nonumber\\
&& \left. \cdot \Big[p_1^{\: 2} + p_2^{\: 2} + 2e\beta \left(s_1 p_2 - s_2 p_1 \right)
+ e^2 \beta^2 \left(s_1^{\: 2} + s_2^{\: 2} \right) \Big] \bigg] \right\}.
\label{eqn:D}
\een}

\section{Details of numerical simulations and results analysis}
\label{sec:particulars}

The obtained complex coupled system of differential equations~\eqref{eqn:P1-2}--\eqref{eqn:D} does not possess an analytic solution and needs to be solved numerically. The~prepared code and tests run to confirm its correctness are presented in appendix~\ref{sec:appendix}.

\begin{figure}[tbp]
\begin{minipage}{0.35\textwidth}
\centering
\includegraphics[width=0.75\textwidth]{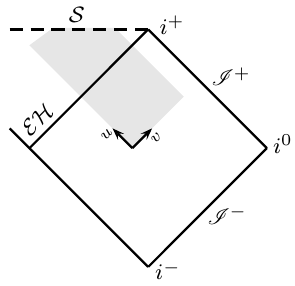}
\end{minipage}
\hfill
\begin{minipage}{0.625\textwidth}
\caption{The computational domain (marked gray) on the background of a~dynamical Schwarzschild spacetime shown on the Carter-Penrose diagram. The~central singularity along \mbox{$r=0$} and the event horizon are denoted as $\mathcal{S}$ and $\mathcal{EH}$, respectively. $\mathscr{I}^\pm$ and $i^\pm$ are null and timelike infinities, respectively, $i^0$ is a~spacelike infinity.}
\label{fig:domain}
\end{minipage}
\end{figure}

The equations describing the dynamics of the considered systems were solved in the bounded region of the $\left(vu\right)$-plane. It~is shown in figure~\ref{fig:domain} on the background of a~dynamical Schwarzschild spacetime, whose Carter-Penrose diagram does not differ from the static case. The~borders of the computational domain were marked for numerical purposes as $0$ and $7.5$ in the $v$-direction, $0$ and $15$ in the $u$-direction in all conducted simulations. The~only arbitrary data of the numerical computations were initial profiles of the evolving fields, posed on the null hypersurface denoted as $u=0$. In~order to describe the behavior of the real and complex scalar fields properly, their initial profiles were of the following Gaussian and trigonometric types, respectively,~\cite{HamadeStewart1996-497,AyalPiran1997-4768,HodPiran1998-1554,OrenPiran2003-044013}:
\ben
h &=& \ah\cdot v^2\cdot e^{-\left(\frac{v-c_1}{c_2}\right)^2},
\label{phi-prof}\\
s &=& \as\cdot \sin^2\left(\pi\frac{v}{v_f}\right)
\cdot\Bigg[\cos\left(\pi\frac{2v}{v_f}\right)+i\cos\left(\pi\frac{2v}{v_f}+\delta\right)\Bigg].
\label{psichi-prof}
\een
The profiles were one-parameter families with amplitudes $\ah$ and $\as$ being the free family parameters. The~amplitudes determine the strength of the gravitational self-interaction of the particular field~\cite{Choptuik1993-9}. The~remaining arbitrary constants were invariable during computations, precisely $c_1=1.3$, $c_2=0.21$ and the parameter of the amount of initial charge $\delta=\frac{\pi}{2}$. The~final value of advanced time was $v_f=2.5$. The~initial conditions are representative for the conducted evolutions, as their outcomes  do not depend of the profiles types provided that they are regular, what means that they result in a~regular spacetime slice at the initial $u=const$ hypersurface. This condition is fulfilled by the above profiles~\eqref{phi-prof} and~\eqref{psichi-prof}.

When the value of the electric coupling constant is not equal to zero, it does not affect the results of the collapse~\cite{BorkowskaRogatkoModerski2011-084007}. It~was confirmed for the investigated cases. Thus $e$ was set as equal to $1$ in all evolutions. The~case of a~vanishing electric coupling constant is beyond the scope of the current research, as it comes down to an analysis of two neutral fields instead of one electrically charged in the spacetime.

The values of the remaining model parameters, i.e., $\alpha$ and $\gamma$ were assigned as follows. The~dilatonic coupling constant was equal to $-\sqrt{3}$, $-1$ and $0$, while $\gamma$ was taken from within the range $\langle -1,1\rangle$. The~values of $\alpha$ corresponded to the dimensionally reduced Kaluza-Klein theory, dilaton gravity and the Einstein-Maxwell theory. The~range of $\gamma$ was consistent with observational constraints (for a~thorough analysis see~\cite{WitekGualtieriPaniSotiriou2019-064035} and references therein).

The spacetime structures resulting from the dynamical evolutions of interest will be presented on Penrose diagrams, which contain contours 
of $r=const$ lines plotted in the $\left(vu\right)$-plane. The~outermost line refers to $r=0$, which in all presented cases will be a~part of a~central singularity. The~lines indicating the vanishing expansion 
\ben
\theta_v\equiv\frac{2}{r}r_{,v}
\label{eqn:expansion}
\een
will be also presented on the diagrams and they will indicate the location of an apparent horizon in the spacetime. On the spacetime diagrams they will be marked as red solid lines. On the remaining plots they will be marked in black.

One of physical quantities which was employed in the interpretation of the obtained results is the quasi-local Hawking mass~\cite{Hawking1968-598}. Its~value calculated for the spherically symmetric spacetime with a~gauge field $A_\mu$ is the following:
\ben
m\left(u,v\right) = \frac{r}{2} \left( 1 + \frac{4fg}{a^2} + \frac{Q^2}{r^2} \right).
\label{haw}
\een
It describes the amount of mass contained within a~sphere of a~radius $r(u,v)$ on a~spacelike hypersurface containing the particular point $(u,v)$. The~mass of a~singular object can be expressed by a~value of the above expression calculated at the event horizon in the region of the emerging spacetime which is not dynamical, that is for large values of advanced time.

The outcomes of the gravitational collapse within the discussed model were also inspected with the use of a~set of local spacetime quantities that may be interpreted as observables related with an observer moving with the evolving matter. The~analysed quantities were energy density $\hat{\rho}$, radial pressure $\hat{p}_{r}$ and pressure anisotropy $\hat{p}_{a} \equiv \hat{p}_{t} - \hat{p}_{r}$. Their precise covariant derivation for any general case can be found in~\cite{NakoniecznaNakonieczny2020-1051}. Moreover, local temperature $T_l = \frac{\kappa_l}{2 \pi}$ with the surface gravity defined as $\kappa_l = a\POu a^{-2}$~\cite{NielsenVisser2006-4637,Nielsen_2008} was also calculated. For the studied gravity-matter model considered in double null coordinates the first three of the above observables are given by relations
\ben
\hat{\rho} &=& e^{2\alpha h}\frac{Q^2}{r^4} + \frac{1}{2 a^2}\left[4\left(x^2 + y^2\right) + e^{2h\left(\alpha+1\right)} \left(p_1^2 + p_2^2 + q_1^2 + q_2^2\right) +\right.\nonumber\\
&&\left. + 2 \beta e^{\left[2h\left(\alpha+1\right)+1\right]} \left(s_1p_2 - s_2p_1\right) + \beta^2 e^{2\left[h\left(\alpha+1\right)+1\right]} \left(s_1^2 + s_2^2\right)\right], \\
\hat{p}_{r} &=& -e^{2\alpha h}\frac{Q^2}{r^4} + \frac{1}{2 a^2}\left[4\left(x^2 + y^2\right) + e^{2h\left(\alpha+1\right)} \left(p_1^2 + p_2^2 + q_1^2 + q_2^2\right) +\right.\nonumber\\
&&\left. + 2 \beta e^{\left[2h\left(\alpha+1\right)+1\right]} \left(s_1p_2 - s_2p_1\right) + \beta^2 e^{2\left[h\left(\alpha+1\right)+1\right]} \left(s_1^2 + s_2^2\right)\right], \\
\hat{p}_{a} &=& 2e^{2\alpha h}\frac{Q^2}{r^4} - \frac{1}{2 a^2}\left\{4\left(x - y\right)^2 + e^{2h\left(\alpha+1\right)} 
\left[\left(p_1 - q_1\right)^2 + \left(p_2 - q_2\right)^2\right] +\right.\nonumber\\
&& \left. + 
  2 \beta e^{\left[2h\left(\alpha+1\right)+1\right]} \left(s_1p_2 - s_1q_2 - s_2p_1 + s_2q_1\right) + 
  \beta^2 e^{2\left[h\left(\alpha+1\right)+1\right]} \left(s_1^2 + s_2^2\right)\right\}.
\een

\section{Dynamical spacetime structures}
\label{sec:struc}

\begin{figure}[tbp]
 \subfigure[][]{\includegraphics[width=0.325\textwidth]{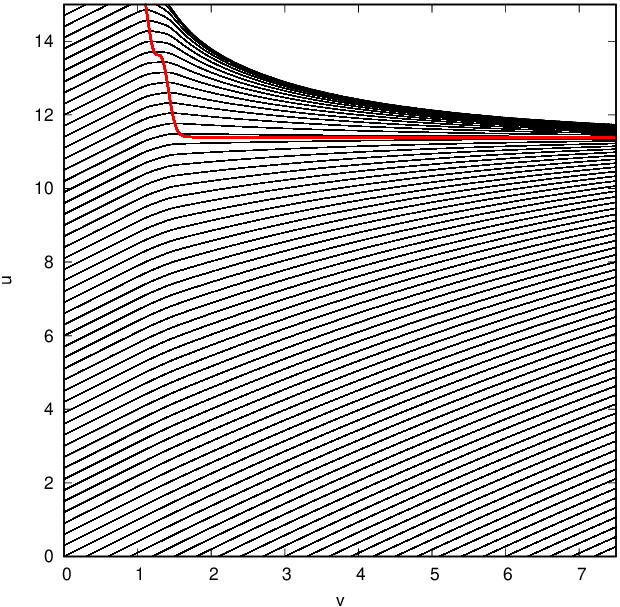}}
 \subfigure[][]{\includegraphics[width=0.325\textwidth]{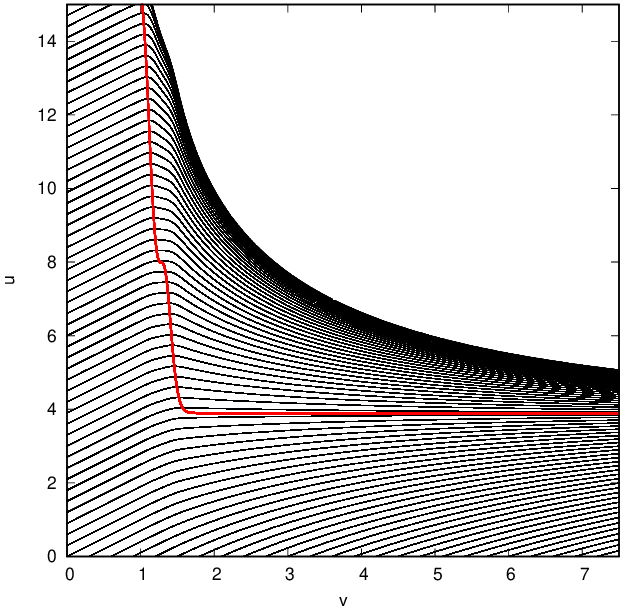}\label{fig:str-1b}}
 \subfigure[][]{\includegraphics[width=0.325\textwidth]{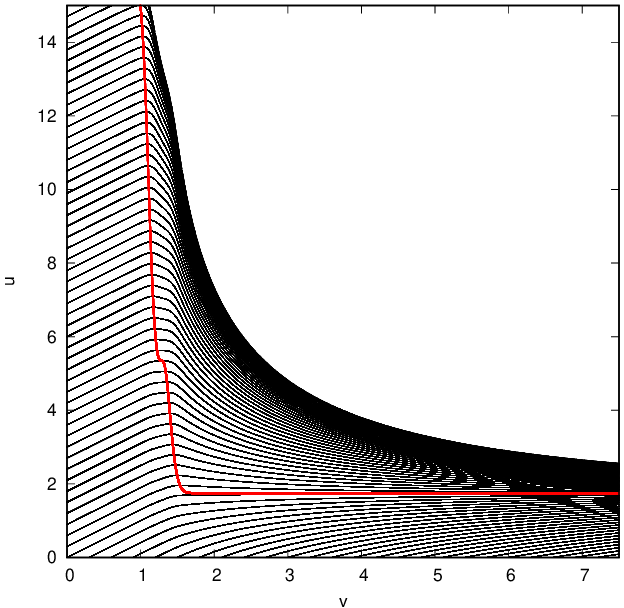}}
\caption{(color online) Penrose diagrams of spacetimes emerging from evolutions with $\alpha=-\sqrt{3}$ and $\gamma=-0.01$. The~field amplitudes $\as=\ah$ were equal to (a)~$0.02$, (b)~$0.04$ and (c)~$0.05$.}
\label{fig:str-1}
\end{figure}

\begin{figure}[tbp]
 \subfigure[][]{\includegraphics[width=0.325\textwidth]{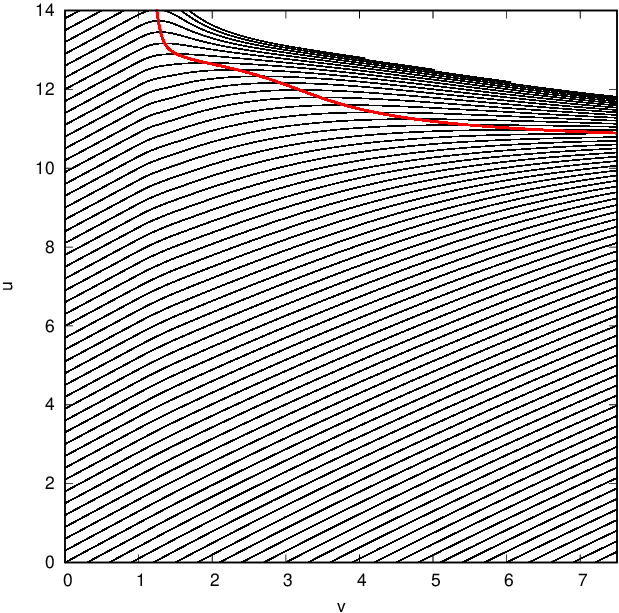}}
 \subfigure[][]{\includegraphics[width=0.325\textwidth]{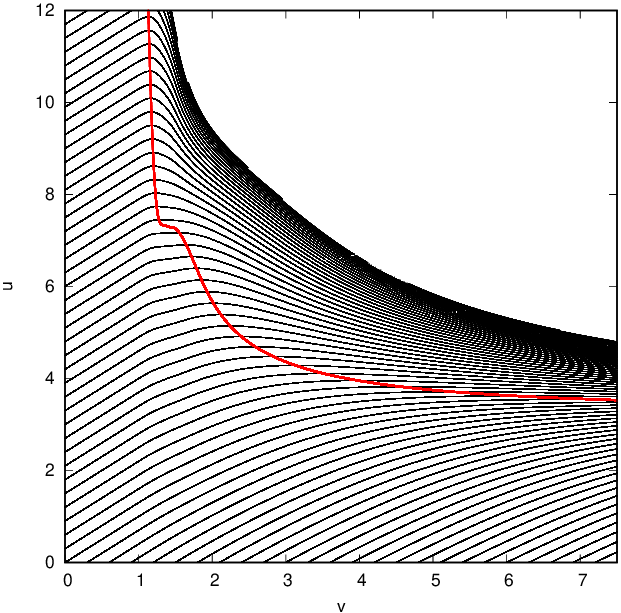}\label{fig:str-2b}}
 \subfigure[][]{\includegraphics[width=0.325\textwidth]{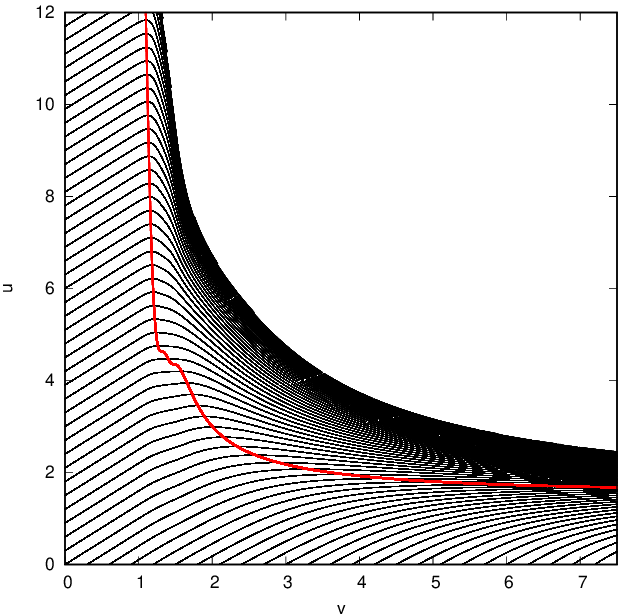}}
\caption{(color online) Penrose diagrams of spacetimes emerging from evolutions with $\alpha=-1$ and $\gamma=-1$. The~field amplitudes $\as=\ah$ were equal to (a)~$0.02$, (b)~$0.04$ and (c)~$0.05$.}
\label{fig:str-2}
\end{figure}

\begin{figure}[tbp]
 \subfigure[][]{\includegraphics[width=0.325\textwidth]{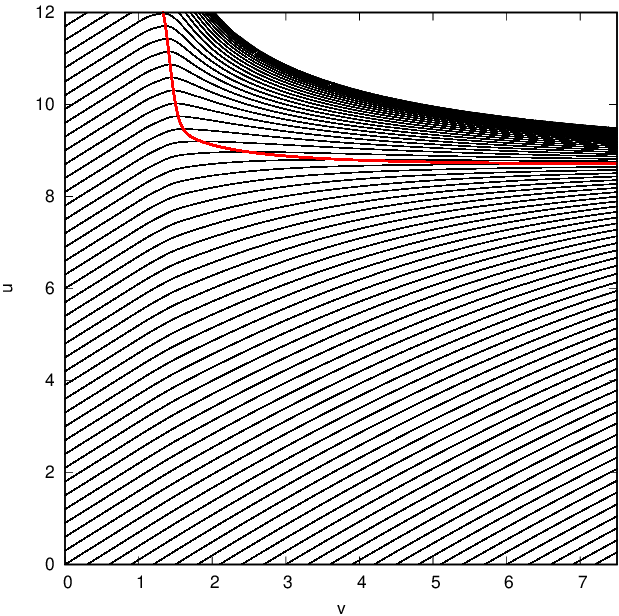}}
 \subfigure[][]{\includegraphics[width=0.325\textwidth]{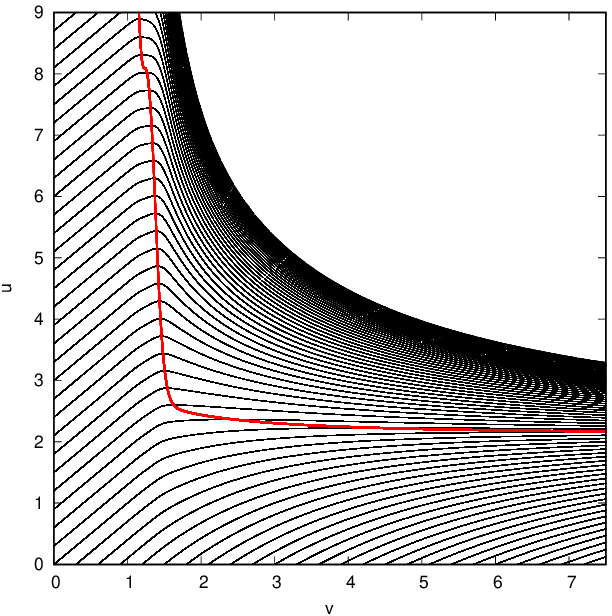}\label{fig:str-3b}}
 \subfigure[][]{\includegraphics[width=0.325\textwidth]{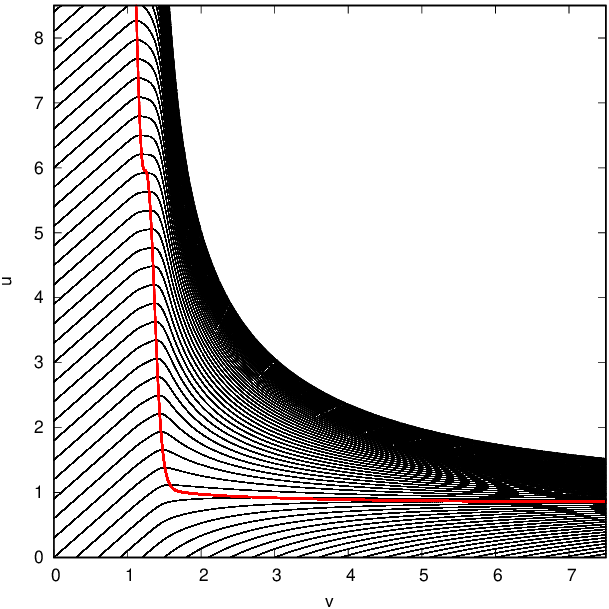}}
\caption{(color online) Penrose diagrams of spacetimes emerging from evolutions with $\alpha=-1$ and $\gamma=1$. The~field amplitudes $\as=\ah$ were equal to (a)~$0.02$, (b)~$0.04$ and (c)~$0.05$.}
\label{fig:str-3}
\end{figure}

\begin{figure}[tbp]
 \subfigure[][]{\includegraphics[width=0.325\textwidth]{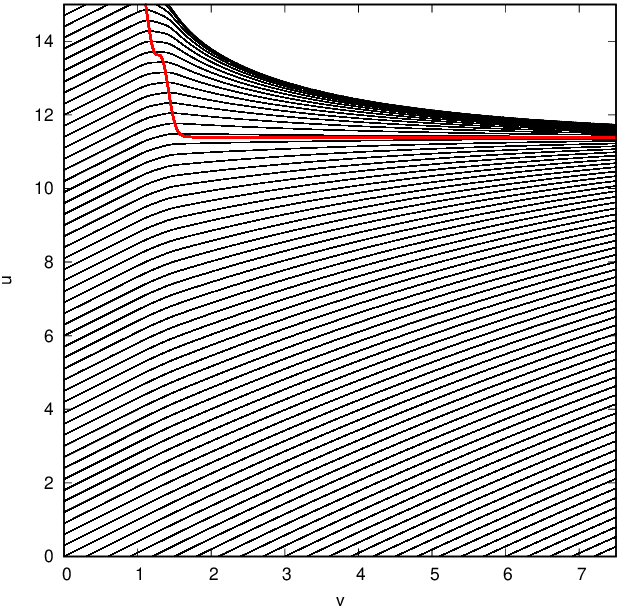}}
 \subfigure[][]{\includegraphics[width=0.325\textwidth]{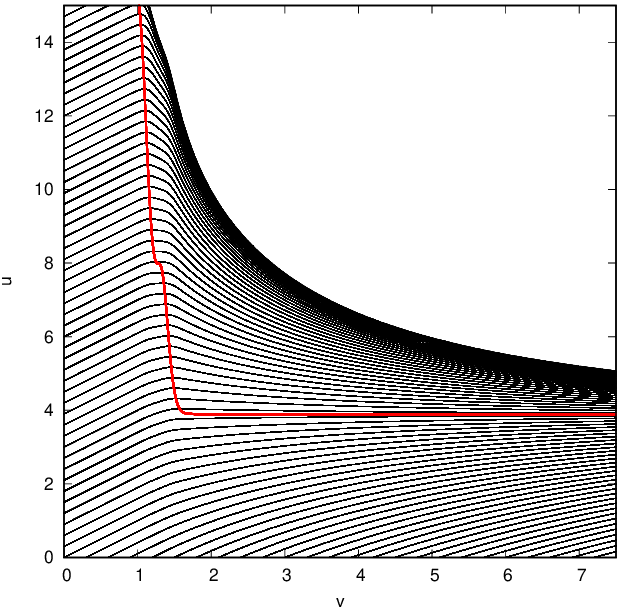}\label{fig:str-4b}}
 \subfigure[][]{\includegraphics[width=0.325\textwidth]{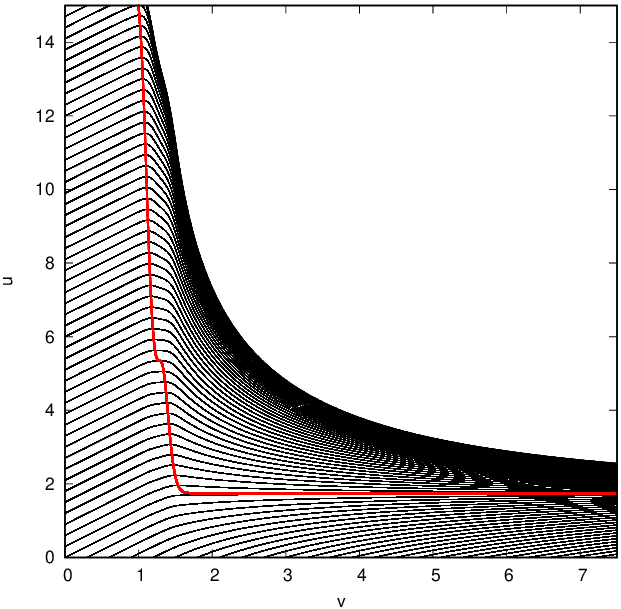}}
\caption{(color online) Penrose diagrams of spacetimes emerging from evolutions with $\alpha=0$ and $\gamma=-0.1$. The~field amplitudes $\as=\ah$ were equal to (a)~$0.02$, (b)~$0.04$ and (c)~$0.05$.}
\label{fig:str-4}
\end{figure}

The spacetimes which emerge during the investigated collapse are either non-singular, for small values of the field self-interaction strengths, or singular, containing in all cases a~dynamical black hole of a~Schwarzschild type, for sufficiently big self-interactions represented by values of field amplitudes. The~singular spacetimes resulting from dynamical evolutions within the model of interest for a~few values of the field amplitudes were presented for several combinations of the model parameters, that is $\alpha=-\sqrt{3}$ and $\gamma=-0.01$, $\alpha=-1$ and $\gamma=-1$, $\alpha=-1$ and $\gamma=1$ and $\alpha=0$ and $\gamma=-0.1$, in figures \ref{fig:str-1}--\ref{fig:str-4}, respectively.

Each resulting dynamical Schwarzschild spacetime contains a~central spacelike singularity located along the $r=0$ line. It~is indicated by the outermost $r=const$ line on the Penrose diagram. The~singularity is surrounded by a~single apparent horizon located along the $r\POv=0$ line, whose behavior divides the whole spacetime into two regions, i.e., a~dynamical one, within which the actual collapse proceeds, and a~non-dynamical one, remaining after the dynamical evolution. The~former corresponds to the range of small values of advanced time, in which the apparent horizon is spacelike and changes its location in the $u$-coordinate. The~latter region refers to the $v$-range of big values, where $v\to\infty$ and the apparent horizon settles along a~null hypersurface of constant retarded time, thus indicating the location of the event horizon in spacetime.

The analysis of the behavior of the apparent horizons in the presented cases allows to draw a~conclusion that the value of $|\gamma|$ influences the size of a~dynamical region within the evolving spacetime. For small absolute values of the coupling the location of the horizon changes significantly within a~small range of the $v$-coordinate and it immediately settles along a~constant-$u$ null hypersurface, where it coincides with the event horizon. When the absolute value of $\gamma$ gets bigger, the region in which the dynamics is observed extends much farther in the $v$-direction and the horizon settles along a~null hypersurface of constant retarded time much later in terms of advanced time. The~transition between the spacelike and null portions of the horizon is less sharp in this case. The~evident dependence of the time of formation of a~black hole and its size on the value of the evolving fields amplitudes will be discussed in detail in section~\ref{sec:char}.

\section{Black hole characteristics}
\label{sec:char}

The examined characteristics of forming black holes were the $u$-locations of the event horizons, radii, masses and electric charges of black holes formed during the gravitational evolutions within the model of interest. They were examined as functions of the model couplings $\alpha\epsilon\langle -\sqrt{3},\sqrt{3}\rangle$ and $\gamma\epsilon\langle -1,1\rangle$, as well as field amplitudes $\as=\ah\epsilon\langle 0.01,0.09\rangle$ for a~selected evolution characterized by the following parameters: $\alpha=-1$, $\gamma=1$, $\as=\ah=0.04$. When the dependence on the specific parameter of the model is presented, the remaining ones are as listed above. In~the case of the dependency on the field self-interaction strengths, several combinations of the model parameters were considered.

\begin{figure}[tbp]
\subfigure[][]{\includegraphics[width=0.47\textwidth]{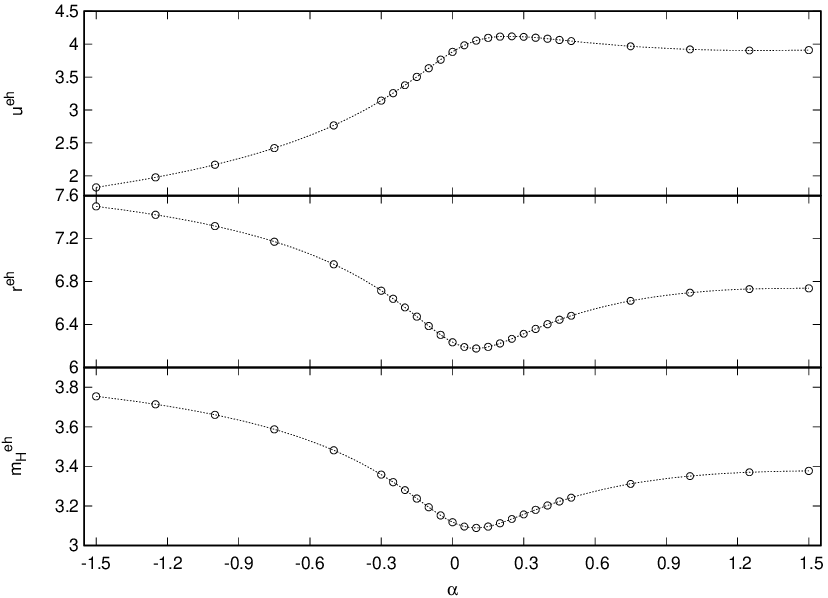}}
\hfill
\subfigure[][]{\includegraphics[width=0.47\textwidth]{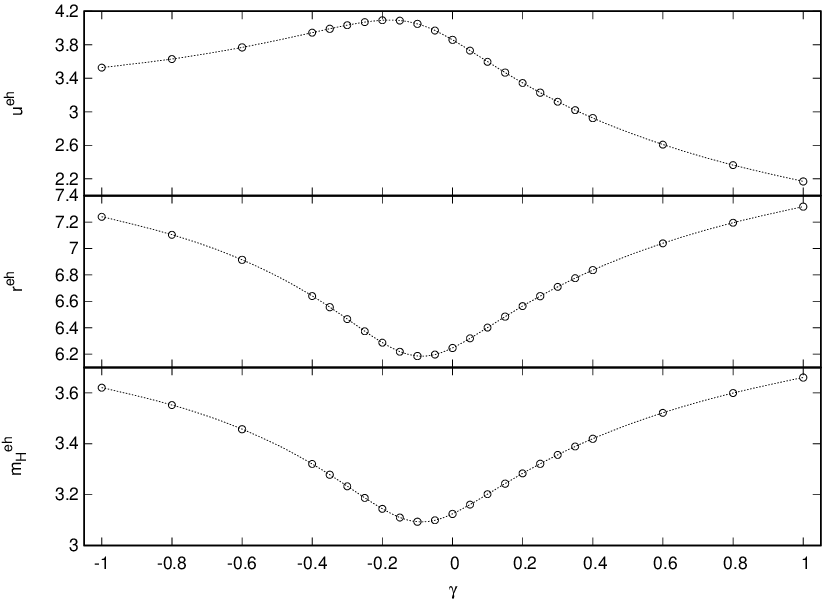}}
\caption{The $u$-locations of the event horizons, $u^{eh}$, radii, $r^{eh}$, and masses, $m_H^{\ eh}$, of black holes formed during the gravitational collapse as functions of (a)~$\alpha$ and (b)~$\gamma$. The~non-varying parameters were $\alpha=-1$ for (b), $\gamma=1$ for (a) and $\as=\ah=0.04$.}
\label{fig:char-urm}
\end{figure}

\begin{figure}[tbp]
\subfigure[][]{\includegraphics[width=0.47\textwidth]{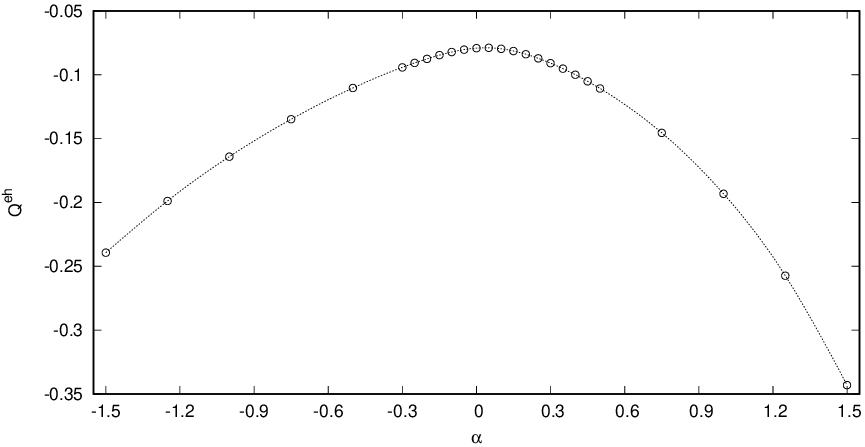}}
\hfill
\subfigure[][]{\includegraphics[width=0.47\textwidth]{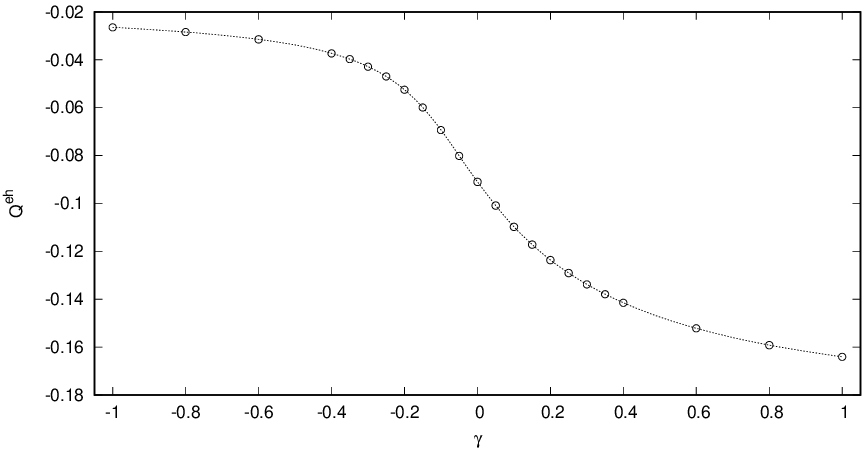}}
\caption{The black hole charge related to the $U(1)$ gauge field, $Q^{eh}$, as function of (a)~$\alpha$ and (b)~$\gamma$, for non-varying parameters as in figure~\ref{fig:char-urm}.}
\label{fig:char-Q}
\end{figure}

The dependencies of the abovementioned nascent black holes characteristics on the model parameters are presented in figures~\ref{fig:char-urm} and~\ref{fig:char-Q}. The~dependencies of the $u$-locations of the event horizons, radii and masses on both $\alpha$ and $\gamma$ are in both cases qualitatively the same, that is there exist a~maximum for $u^{eh}$ at $\alpha=0.25$ and $\gamma=-0.2$ and minima at $\alpha=0.1$ and $\gamma=-0.1$ for the remaining characteristics. The~black hole electric charge increases for small values of the dilatonic coupling constant up to an extremum $\alpha=0.05$ and decreases for its larger values. $Q^{eh}$ decreases monotonically with an increasing $\gamma$.

\begin{figure}[tbp]
\subfigure[][]{\includegraphics[width=0.47\textwidth]{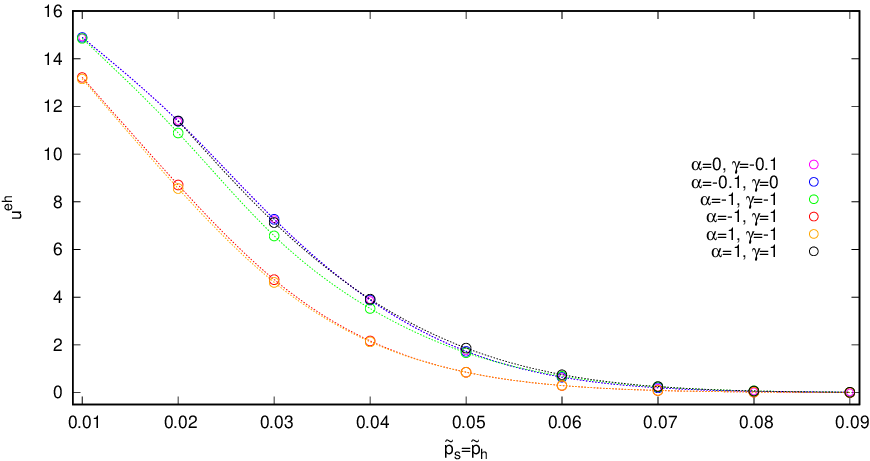}}
\hfill
\subfigure[][]{\includegraphics[width=0.47\textwidth]{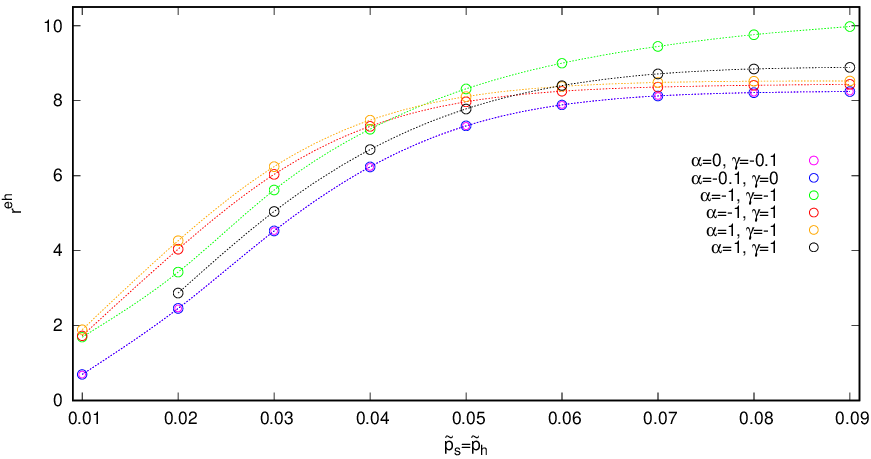}}\\
\subfigure[][]{\includegraphics[width=0.47\textwidth]{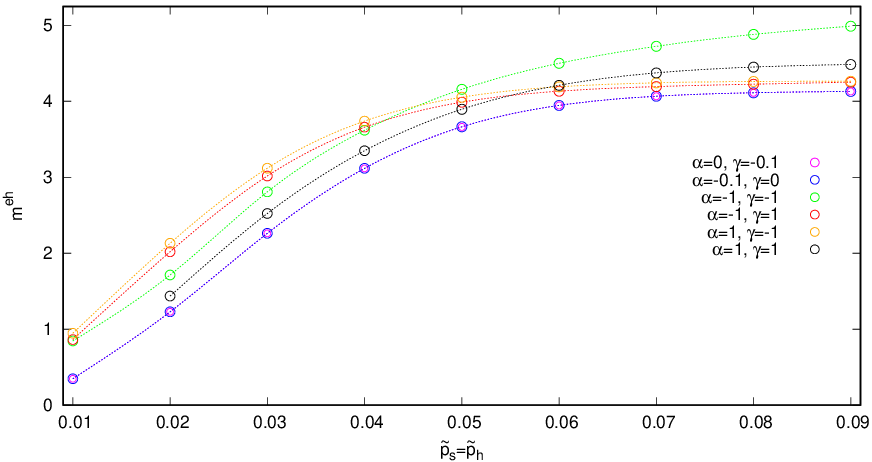}}
\hfill
\subfigure[][]{\includegraphics[width=0.47\textwidth]{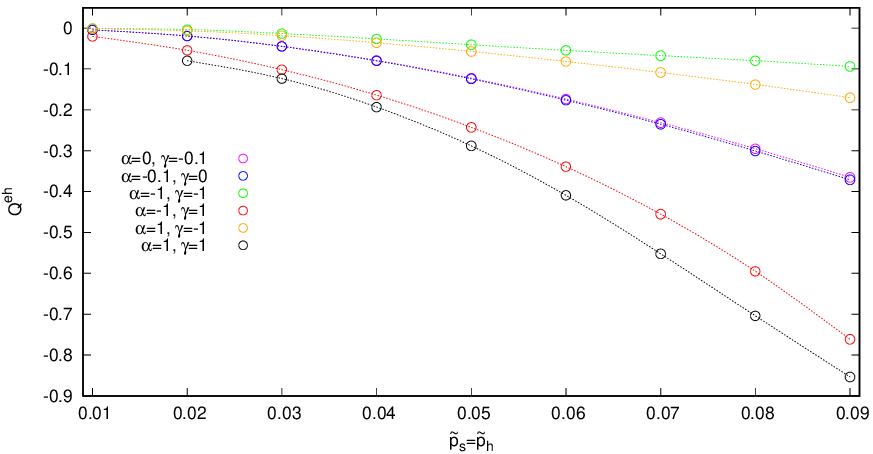}}
\caption{(color online) The~(a)~$u$-locations of the event horizons, $u^{eh}$, (b)~radii, $r^{eh}$, (c)~masses, $m_H^{\ eh}$, and (d)~charges related to the $U(1)$ gauge field, $Q^{eh}$, of black holes formed during the gravitational collapse as functions of the field amplitudes, $\as$ and $\ah$, for several combinations of values of the parameters $\alpha$ and $\gamma$.}
\label{fig:char-fa}
\end{figure}

Figure~\ref{fig:char-fa} presents the discussed black hole characteristics as functions of the collapsing fields amplitudes for several combinations of the couplings present within the studied model. In~all cases, for increasing field self-interaction strengths the changes of the black hole characteristics are monotonic. The~$u$-locations of the event horizons and black hole electric charges decrease, while radii and masses of the forming singular objects increase with increasing $\as=\ah$. The~changes of $u^{eh}$, $r^{eh}$ and $m_H^{\ eh}$ become smaller with increasing field amplitudes. An opposite tendency is observed for $Q^{eh}$.

\section{Observables and fields}
\label{sec:obs}

The $\left(vu\right)$-distributions of observables presented in section~\ref{sec:particulars} and the evolving scalar fields, along with the relation between local temperature along the apparent horizon and the $v$-coordinate will be shown and discussed for selected spacetimes, the structures of which were shown in section~\ref{sec:struc}.

\begin{figure}[tbp]
\centering
\subfigure[][]{\includegraphics[width=0.325\textwidth]{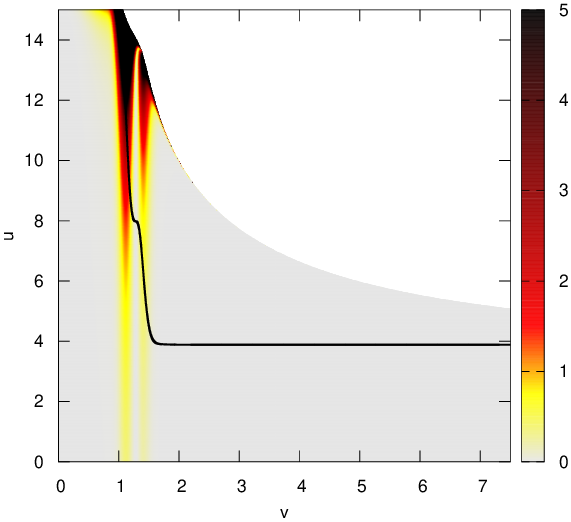}}
\subfigure[][]{\includegraphics[width=0.325\textwidth]{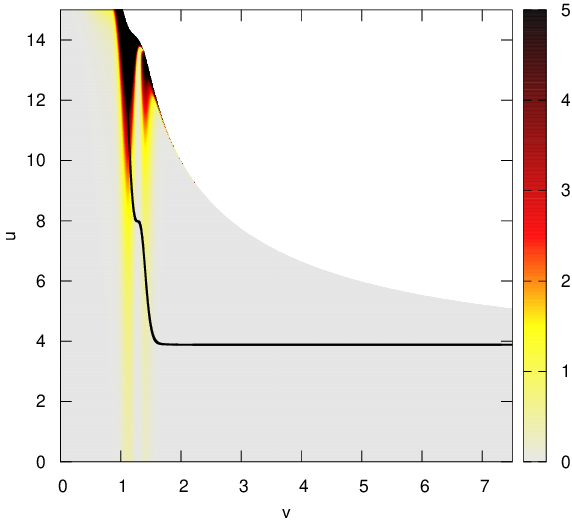}}
\subfigure[][]{\includegraphics[width=0.325\textwidth]{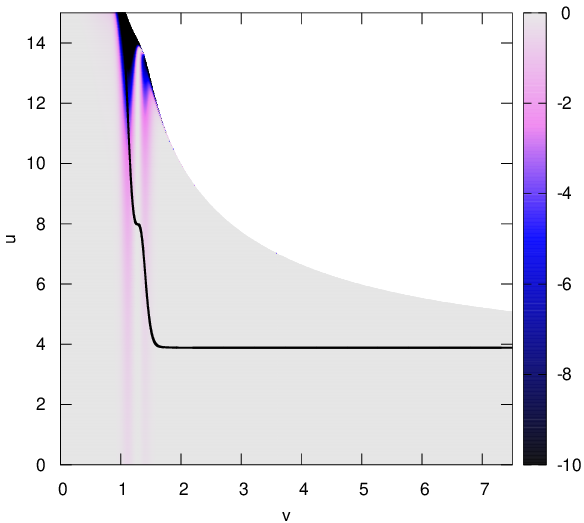}}\\
\begin{minipage}{0.5\textwidth}
\subfigure[][]{\includegraphics[width=0.95\textwidth]{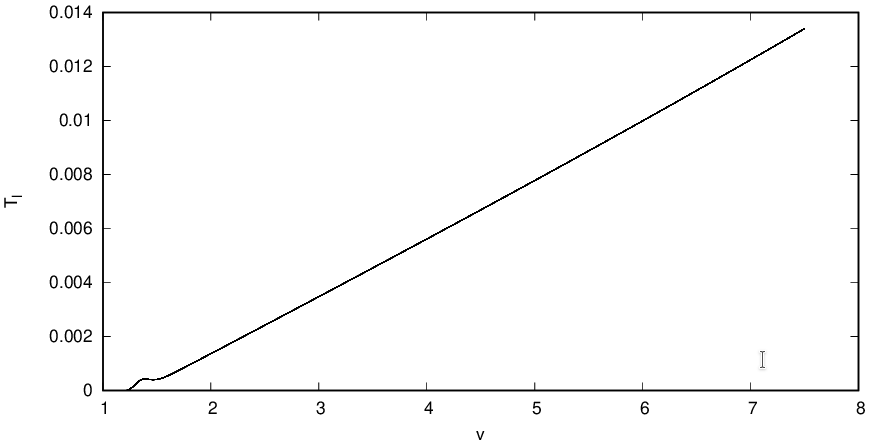}\label{fig:coupl-obs-1ltah}}
\end{minipage}
\begin{minipage}{0.475\textwidth}
\caption{(color online) The~$\left(vu\right)$-distribution of (a)~energy density, $\hat{\rho}$, (b)~radial pressure, $\hat{p}_{r}$, and (c)~pressure anisotropy, $\hat{p}_{a}$, and (d)~local temperature along the black hole apparent horizon, $T_l$, as a~function of advanced time for a~dynamical evolution characterized by parameters $\alpha=-\sqrt{3}$, $\gamma=-0.01$ and $\as=\ah=0.04$ (the same as in figure~\ref{fig:str-1b}).}
\label{fig:coupl-obs-1}
\end{minipage}
\end{figure}

\begin{figure}[tbp]
\centering
\subfigure[][]{\includegraphics[width=0.325\textwidth]{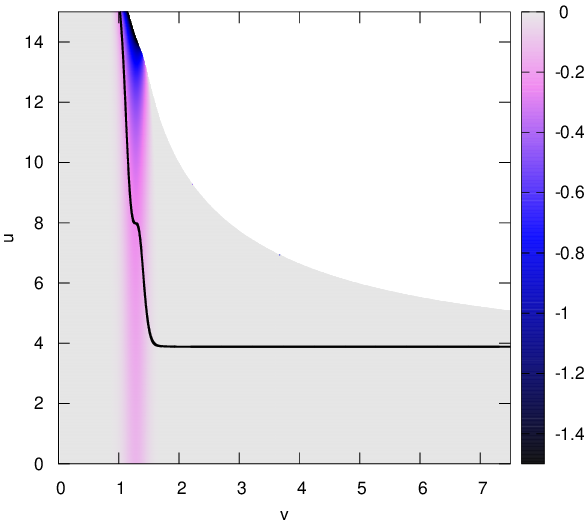}}
\hspace{1cm}
\subfigure[][]{\includegraphics[width=0.325\textwidth]{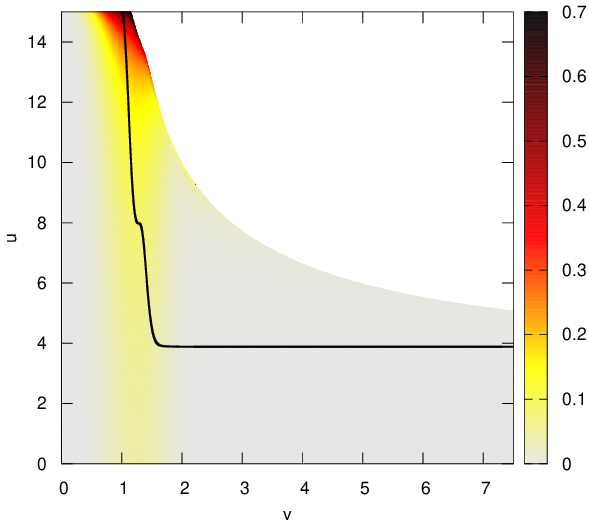}}
\caption{(color online) The~$\left(vu\right)$-distribution of (a)~the neutral scalar field, $h$, and (b)~the moduli of the complex scalar field, $|s|$, for the same parameters and field amplitudes as in figure~\ref{fig:coupl-obs-1}.}
\label{fig:coupl-fie-1}
\end{figure}

Figures~\ref{fig:coupl-obs-1} and~\ref{fig:coupl-fie-1} present the considered spacetime distributions for the spacetime whose structure was shown in figure~\ref{fig:str-1b}, with the model parameters $\alpha=-\sqrt{3}$ and $\gamma=-0.01$. The~highest absolute values of the discussed observables as well as both the neutral and the moduli of the complex scalar field functions are located along a~constant-$v$ null direction, which is a~direction of peaks of initially imposed field functions propagation. A~considerable increase in their values appears as the central singularity is approached. This increase is observed only within the range of advanced time, in which the highest field values were imposed initially. The~energy density, radial pressure and the moduli of the complex scalar field are positive within the whole integration domain. On the contrary, the pressure anisotropy and the neutral scalar field function are negative there. The~black hole local temperature calculated along the apparent horizon is positive and it increases with advanced time. The~changes are not monotonic. It~increases for small values of the $v$-coordinate, reaches a~maximum within the $v$-range, where there exists a~small inclination of the apparent horizon and then, after a~slight decrease, it increases monotonically in a~nearly linear manner.

\begin{figure}[tbp]
\centering
\subfigure[][]{\includegraphics[width=0.325\textwidth]{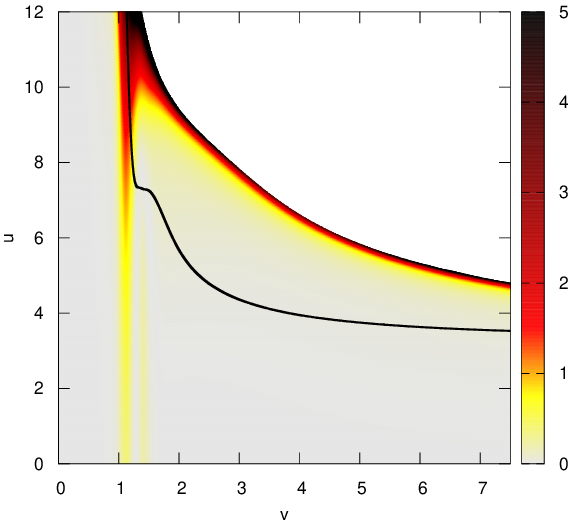}}
\subfigure[][]{\includegraphics[width=0.325\textwidth]{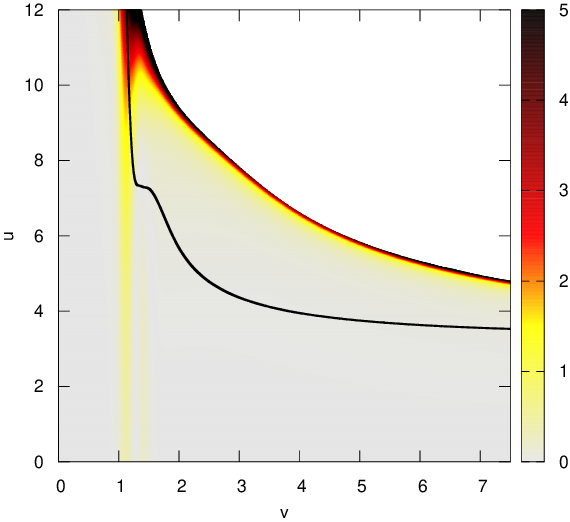}}
\subfigure[][]{\includegraphics[width=0.325\textwidth]{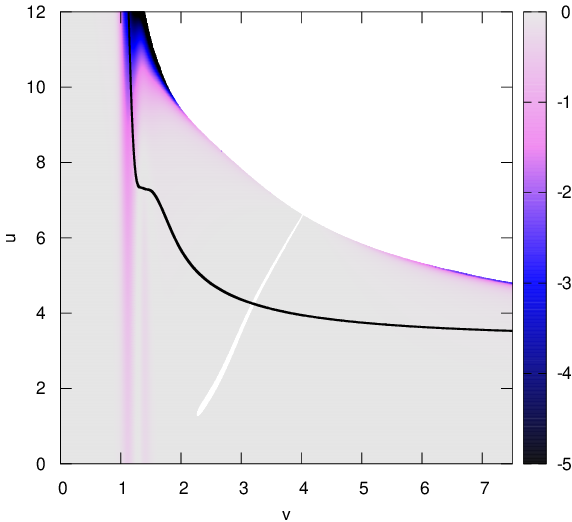}}\\
\begin{minipage}{0.5\textwidth}
\subfigure[][]{\includegraphics[width=0.95\textwidth]{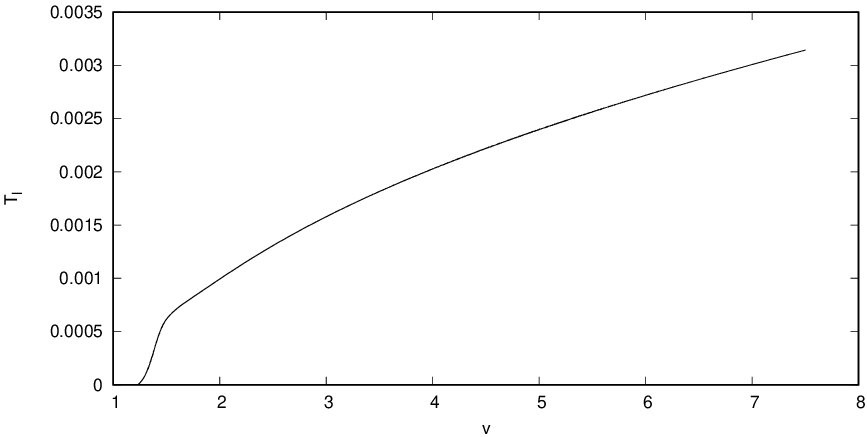}}
\end{minipage}
\begin{minipage}{0.475\textwidth}
\caption{(color online) The~$\left(vu\right)$-distribution of (a)~energy density, $\hat{\rho}$, (b)~radial pressure, $\hat{p}_{r}$, and (c)~pressure anisotropy, $\hat{p}_{a}$, and (d)~local temperature along the black hole apparent horizon, $T_l$, as a~function of advanced time for a~dynamical evolution characterized by parameters $\alpha=-1$, $\gamma=-1$ and $\as=\ah=0.04$ (the same as in figure~\ref{fig:str-2b}).}
\label{fig:coupl-obs-2}
\end{minipage}
\end{figure}

\begin{figure}[tbp]
\centering
\subfigure[][]{\includegraphics[width=0.325\textwidth]{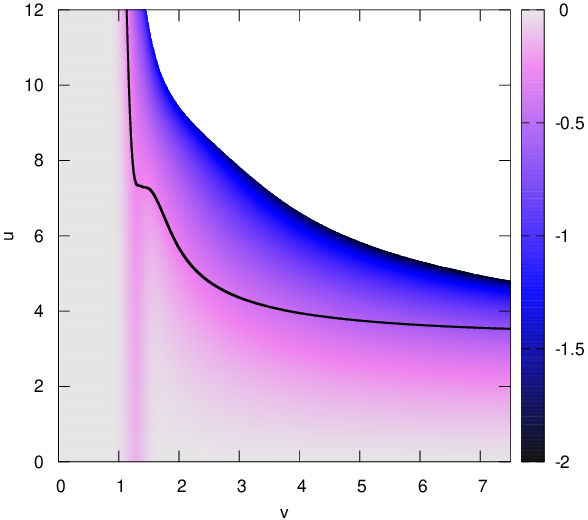}\label{fig:coupl-fie-2a}}
\hspace{1cm}
\subfigure[][]{\includegraphics[width=0.325\textwidth]{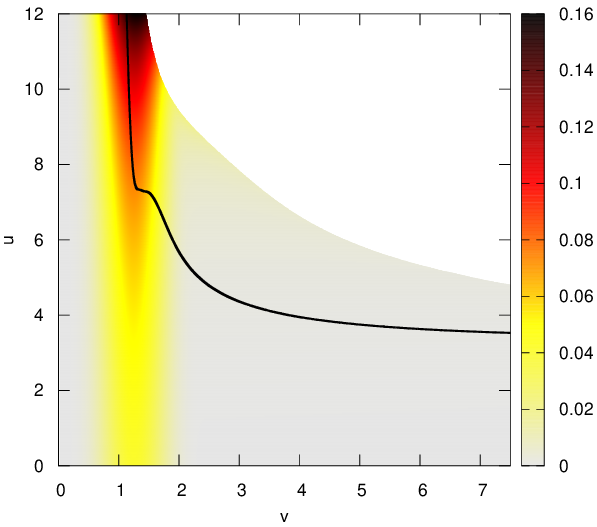}}
\caption{(color online) The~$\left(vu\right)$-distribution of (a)~the neutral scalar field, $h$, and (b)~the moduli of the complex scalar field, $|s|$, for the same parameters and field amplitudes as in figure~\ref{fig:coupl-obs-2}.}
\label{fig:coupl-fie-2}
\end{figure}

The distributions of observables and field functions resulting from the evolution leading to the spacetime with the structure shown in figure~\ref{fig:str-2b}, obtained with $\alpha=-1$ and $\gamma=-1$, are presented in figures~\ref{fig:coupl-obs-2} and~\ref{fig:coupl-fie-2}, respectively. As~in the case discussed above, the energy density and the moduli of the complex scalar field are positive, while the pressure anisotropy and the neutral scalar field function are negative within the whole domain of integration. 
What distinguishes this case from the other discussed is the fact that non-zero energy density and radial pressure persist also for large values of advanced time. Moreover, pressure anisotropy is also non-zero for large $v$. This implies that there is a~persisting non-trivial matter distribution around the central singularity. This observation finds its confirmation in figure~\ref{fig:coupl-fie-2a}. The~dilaton field does not collapse completely but instead attains a~non-zero value even outside the black hole event horizon. This may indicate that in this case a~formation of a~hairy black hole is observed. 
The local temperature in the examined case is positive and increases monotonically with advanced time along the apparent horizon with a~slight inclination in the region, where an inclination of the apparent horizon is also visible. Unlike the case depicted in figure~\ref{fig:coupl-obs-1ltah}, the late-lime increase is not linear.

\begin{figure}[tbp]
\centering
\subfigure[][]{\includegraphics[width=0.325\textwidth]{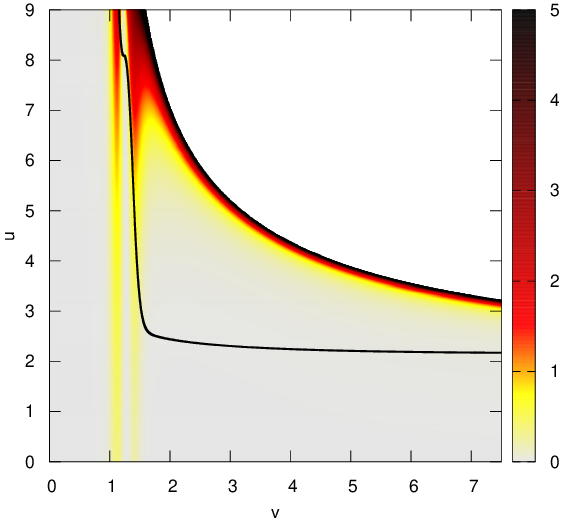}}
\subfigure[][]{\includegraphics[width=0.325\textwidth]{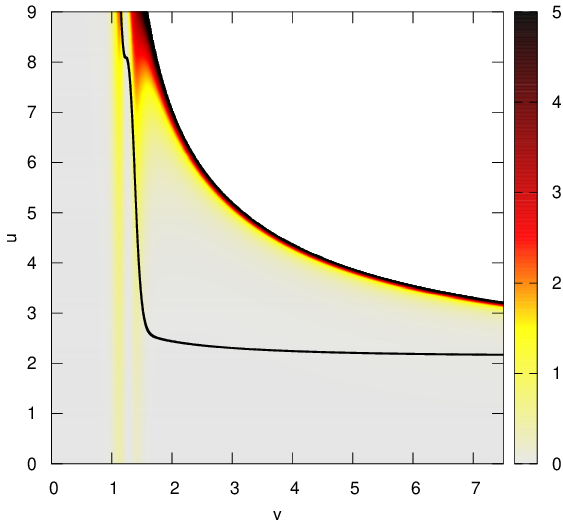}}
\subfigure[][]{\includegraphics[width=0.325\textwidth]{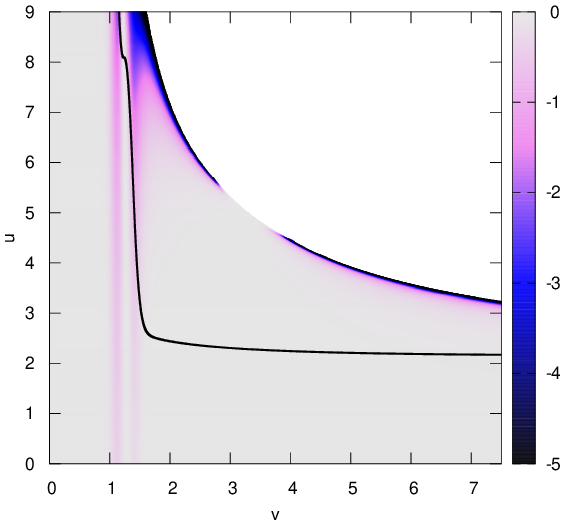}}\\
\begin{minipage}{0.5\textwidth}
\subfigure[][]{\includegraphics[width=0.95\textwidth]{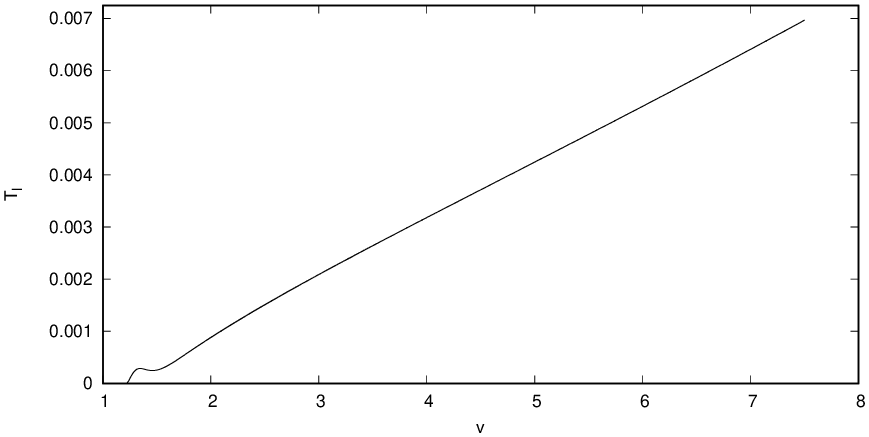}}
\end{minipage}
\begin{minipage}{0.475\textwidth}
\caption{(color online) The~$\left(vu\right)$-distribution of (a)~energy density, $\hat{\rho}$, (b)~radial pressure, $\hat{p}_{r}$, and (c)~pressure anisotropy, $\hat{p}_{a}$, and (d)~local temperature along the black hole apparent horizon, $T_l$, as a~function of advanced time for a~dynamical evolution characterized by parameters $\alpha=-1$, $\gamma=1$ and $\as=\ah=0.04$ (the same as in figure~\ref{fig:str-3b}).}
\label{fig:coupl-obs-3}
\end{minipage}
\end{figure}

\begin{figure}[tbp]
\centering
\subfigure[][]{\includegraphics[width=0.325\textwidth]{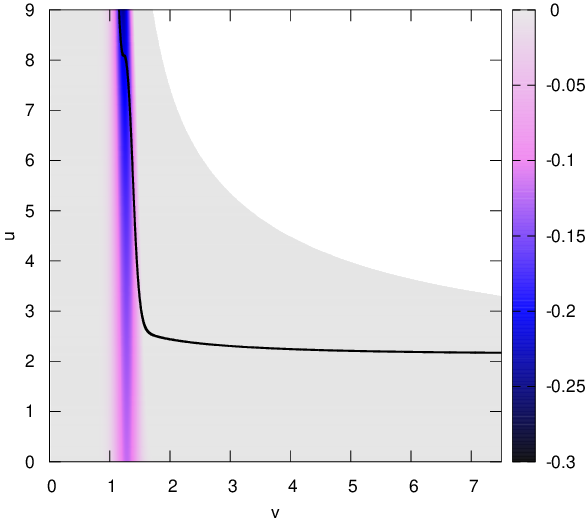}}
\hspace{1cm}
\subfigure[][]{\includegraphics[width=0.325\textwidth]{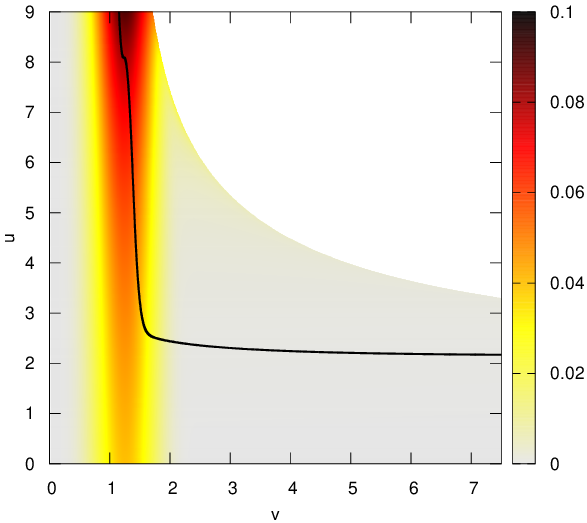}}
\caption{(color online) The~$\left(vu\right)$-distribution of (a)~the neutral scalar field, $h$, and (b)~the moduli of the complex scalar field, $|s|$, for the same parameters and field amplitudes as in figure~\ref{fig:coupl-obs-3}.}
\label{fig:coupl-fie-3}
\end{figure}

The distributions of the discussed quantities, that is observables and field functions, for the evolution with the emerging spacetime structure presented in figure~\ref{fig:str-3b}, resulting from the evolution with $\alpha=-1$ and $\gamma=1$, are shown in figures~\ref{fig:coupl-obs-3} and~\ref{fig:coupl-fie-3}. The~signs of the particular quantities are the same as in the cases presented above. An increase of absolute values of the observables is visible in the region corresponding to advanced time where the field values was the highest initially. The~energy density, radial pressure and pressure anisotropy also display an increase in values nearby the central singularity, as in the latter of the above cases. The~local temperature along the black hole apparent horizon behaves like in the first case above, namely it increases for small values of $v$, then, after reaching a~maximum it slightly decreases and the decrease turns into the late-time nearly linear increase as $v\to\infty$.

\begin{figure}[tbp]
\centering
\subfigure[][]{\includegraphics[width=0.325\textwidth]{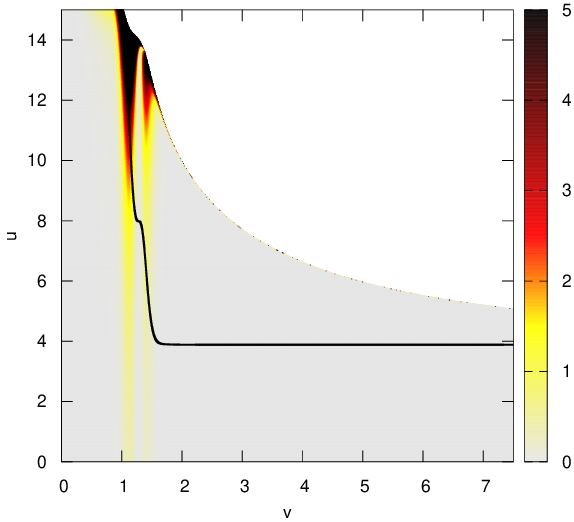}}
\subfigure[][]{\includegraphics[width=0.325\textwidth]{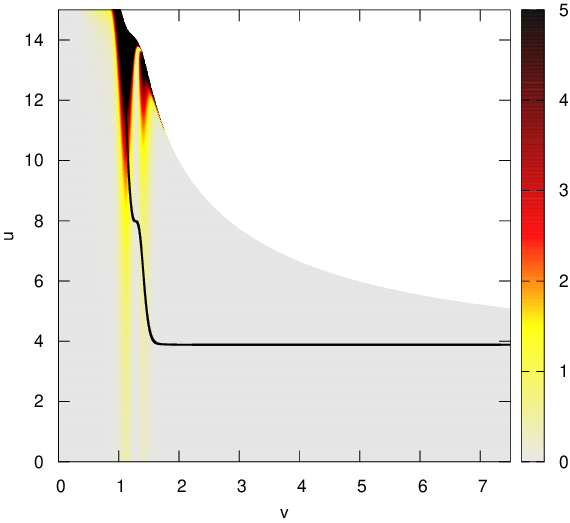}}
\subfigure[][]{\includegraphics[width=0.325\textwidth]{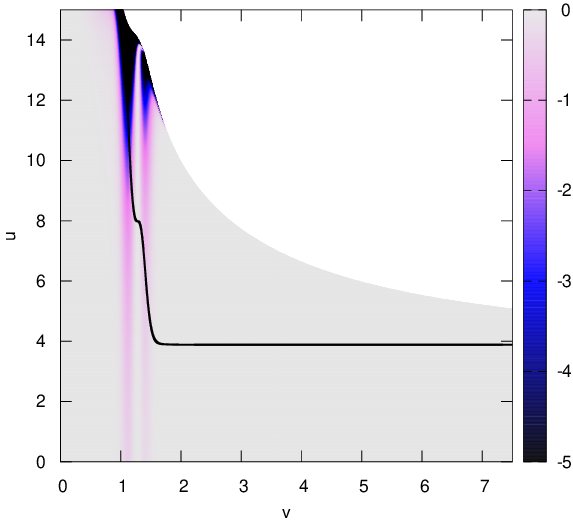}}\\
\begin{minipage}{0.5\textwidth}
\subfigure[][]{\includegraphics[width=0.95\textwidth]{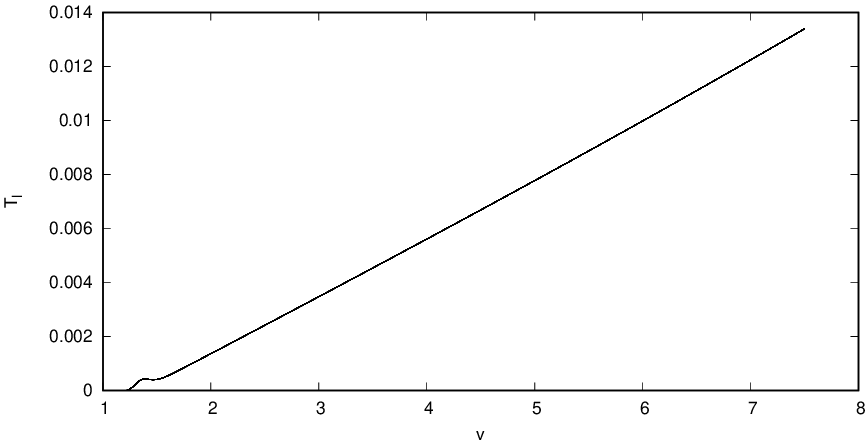}}
\end{minipage}
\begin{minipage}{0.475\textwidth}
\caption{(color online) The~$\left(vu\right)$-distribution of (a)~energy density, $\hat{\rho}$, (b)~radial pressure, $\hat{p}_{r}$, and (c)~pressure anisotropy, $\hat{p}_{a}$, and (d)~local temperature along the black hole apparent horizon, $T_l$, as a~function of advanced time for a~dynamical evolution characterized by parameters $\alpha=0$, $\gamma=-0.1$ and $\as=\ah=0.04$ (the same as in figure~\ref{fig:str-4b}).}
\label{fig:coupl-obs-4}
\end{minipage}
\end{figure}

\begin{figure}[tbp]
\centering
\subfigure[][]{\includegraphics[width=0.325\textwidth]{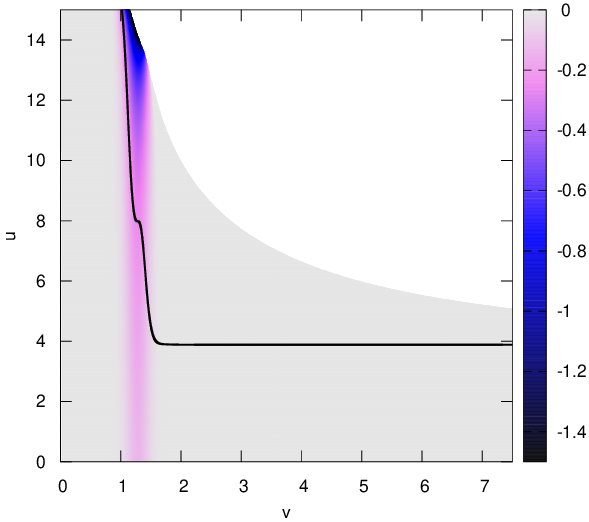}}
\hspace{1cm}
\subfigure[][]{\includegraphics[width=0.325\textwidth]{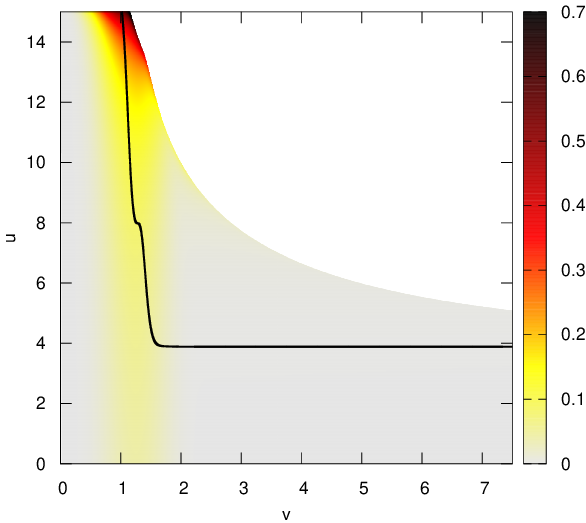}}
\caption{(color online) The~$\left(vu\right)$-distribution of (a)~the neutral scalar field, $h$, and (b)~the moduli of the complex scalar field, $|s|$, for the same parameters and field amplitudes as in figure~\ref{fig:coupl-obs-4}.}
\label{fig:coupl-fie-4}
\end{figure}

The spacetime distribution of the analysed observables and fields for the evolution with $\alpha=0$ and $\gamma=-0.1$, with the structure of spacetime shown in figure~\ref{fig:str-4b}, presented in figures~\ref{fig:coupl-obs-4} and~\ref{fig:coupl-fie-4}, is analogous to the one presented above for the case of $\alpha=-\sqrt{3}$ and $\gamma=-0.01$. The~energy density, radial pressure and the moduli of the complex scalar field are positive, while the pressure anisotropy and the neutral scalar field function are negative within the dynamical spacetime region covered by numerical calculations. A~sole increase in absolute values in all the distributions is visible along the null $v=const$ direction of propagation of the highest initial values of the imposed field functions. The~changes in values of the local temperature calculated along the black hole apparent horizon tend to a~late-time increase close to linear, after reaching a~shallow extrema, precisely a~maximum and a~minimum for small values of advanced time in the range where an inclination of the apparent horizon appears.

\section{Conclusions}
\label{sec:conclusions}

The course and outcomes of gravitational evolutions involving electric charge in a~decoupling limit of the dilatonic Gauss--Bonnet gravity was investigated. The~system consisted of two scalar fields, namely a~neutral scalar dilaton field and a~complex scalar field coupled with a~$U(1)$ gauge field, i.e., electrically charged. The~dilaton was coupled both with the matter and Gauss--Bonnet sectors of the theory.

The course of the dynamical collapse was traced numerically. The~formation of dynamical spacetimes and emerging black holes was observed and analysed via description of spacetime structures, inspection of the black hole characteristics and values of gravitational observables and evolving fields within the forming spacetimes.

The collapse resulted in either non-singular spacetimes formed for small self-interaction strengths of the fields, that is small initial values of their amplitudes, or singular spacetimes containing black holes. The~particular spacetime contained a~central spacelike singularity along $r=0$ surrounded by a~single apparent horizon, which was spacelike in the dynamical region, i.e., for small values of advanced time, and became null as $v\to\infty$. Disregarding the values of the model parameters $\alpha$ and $\gamma$, the black holes were of a~Schwarzschild type, despite the fact that the evolving scalar field was electrically charged. Such a~situation was observed in the case of a~dynamical gravitational collapse of an electrically charged scalar field in dilaton gravity, but only for non-zero values of the dilatonic coupling constant~\cite{BorkowskaRogatkoModerski2011-084007}. Taking this into account, a~conclusion that the Gauss--Bonnet term in the gravitational sector suppresses the tendency to form a~Reissner-Nordstr\"{o}m spacetime with an inner Cauchy horizon, whose appearance is characteristic for evolutions involving electric charge, can be drawn. The~absolute value of the parameter $\gamma$ influenced the $v$-range of the dynamical spacetime region, precisely the range of advanced time in which gravitational dynamics was observed was narrower for smaller values of $|\gamma|$.

The dependencies of the $u$-locations of the event horizons, radii and masses of black holes emerging from the investigated process on $\alpha$ and $\gamma$ are qualitatively the same. As~values of the model parameters increase, the black holes form later in terms of retarded time and both their radii and masses become smaller, up to an extremum, in which the tendencies reverse. Similar dependencies were observed in the case of non-minimal scalar--gravity couplings during dynamics investigations within the theory involving Higgs and dark matter sectors~\cite{NakoniecznaNakonieczny2020-1051}. The~only difference between the dependencies on the model parameters is that the changes are more significant for small values of $\alpha$ and large values of $\gamma$. The~dependence of the values of black hole $U(1)$ charge on $\alpha$ and $\gamma$ do not overlap the ones describe above. In~the case of the dilatonic coupling constant, $Q^{eh}$ increases significantly for its small values and decreases also significantly after reaching a~maximum. In~the case of the Gauss--Bonnet coupling, the black hole electric charge decreases monotonically within the whole parameter range.

The changes of the $u$-locations of the event horizons, radii, masses and electric charges of the nascent black holes with increasing self-interaction strengths of the evolving scalar fields are monotonic and qualitatively the same for various combinations of values of the model parameters $\alpha$ and $\gamma$. The~features $u^{eh}$ and $Q^{eh}$ increase, while $r^{eh}$ and $m_H^{\ eh}$ decrease with the parameters. The~changes are more significant for the electric charge in comparison to the remaining black hole characteristics.

In all the investigated cases an increase of the energy density, radial pressure, pressure anisotropy and values of the collapsing scalar fields was observed along a~null direction of propagation of the maxima of initially imposed field profiles in spacetime. For large absolute values of the $\gamma$ parameter another increase in values of the quantities measured by an observer moving with the collapsing matter was visible in a~close vicinity of the emerging singularity also for large values of advanced time. This implies a~persisting non-trivial matter distribution around the central singularity. Additionally, an observation that the~dilaton field possesses non-zero values outside the black hole event horizon may indicate a~formation of a~hairy black hole in this case. The~local temperature calculated along the apparent horizon of the emerging black holes shows a~late-time monotonic increase for all the investigated cases. Within dynamical spacetime regions, where inclinations of the horizons appear, the changes of the values of local temperature are not monotonic and extrema are observed.

The ultimate aim of the undertaken research is to investigate gravitational evolution of scalar fields in the full version of the EdGB theory in double null coordinates. These coordinates cover the whole spacetime, that is both the exterior and interior of objects that may arise in dynamically formed spacetimes. The~presented analysis within the truncated version of the underlying theory is a~first step towards calculations in the full EdGB. Its~equations require a~more sophisticated treatment, as they are coupled in a~complicated way. The~issue has been only partially resolved by introducing a~shift symmetry in the case of coordinates, which do not penetrate emerging horizons, i.e., when only the exterior of the forming objects is possible to be thus dealt with~\cite{RipleyPretorius2019-134001}.

\appendix
\section{Numerical computations}
\label{sec:appendix}

\subsection*{Algorithm setup}

The evolution of the studied physical system is described by equations~\eqref{eqn:P1-2}--\eqref{eqn:D}. It~was resolved numerically. The~set of equations of motion involves quantities $d$, $q_1$, $q_2$, $y$, $s_1$, $s_2$, $h$, $a$, $p_1$, $p_2$, $x$, $r$, $f$, $g$, $Q$, $\beta$. Each function depends on two null coordinates, namely advanced and retarded times. The~dynamics of $d$, $q_1$, $q_2$ and $y$ was followed along $u$ in line with the equations $E4$, $S_{_{\left(Re\right)}}$, $S_{_{\left(Im\right)}}$ and $D$, respectively. The~remaining quantities, $s_1$, $s_2$, $h$, $a$, $p_1$, $p_2$, $x$, $r$, $f$, $g$, $Q$ and $\beta$, evolved along $v$ according to the respective equations $P6$, $P7$, $P2$, $S_{_{\left(Re\right)}}$, $S_{_{\left(Im\right)}}$, $D$, $P4$, $E3$, $E2$, $M2$ and $M1$.

The system of evolution equations was solved in a~bounded region of the $\left(vu\right)$-plane presented in figure~\ref{fig:domain} in section~\ref{sec:particulars}. An arbitrary null hypersurface of constant retarded time was taken as an initial data surface. The~boundary conditions were posed on a~hypersurface of constant advanced time. The~two surfaces were marked as $u=0$ and $v=0$, respectively, for computational purposes. 

Initial conditions are arbitrary profiles of the fields functions, $s_1$, $s_2$ and $h$, which were posed according to~\eqref{psichi-prof} and~\eqref{phi-prof}. The~initial values along $v$ of $q_1$, $q_2$ and $y$ were calculated analytically using the relations $P6$ and $P7$. In~the employed setup the distribution of matter is shell--shaped, hence the boundary is unaffected by it and the field functions $s_1$, $s_2$ and $h$ vanish there. The~boundary values of $q_1$, $q_2$ and $y$ were obtained through integration of equations $S_{_{\left(Re\right)}}$, $S_{_{\left(Im\right)}}$ and $D$, respectively.

A gauge freedom to choose initial and boundary profiles of the $r$ function remains within the investigated setup. $r\left(0,0\right)$ was chosen to be equal to $7.5$ for computational purposes. Initial and boundary values of $g$ and $f$, respectively, determine the distances between the null lines and were chosen to be constant, that is $g\left(0,v\right)=\frac{1}{2}$ and $f\left(v,0\right)=-\frac{1}{2}$. These values are justified by the fact that mass~\eqref{haw} should vanish at the central point $\left(0,0\right)$. The~$r$ values on the initial null segment were obtained using the relation $P4$ and along the boundary with the equation $P3$. Initial and boundary profiles of $f$ and $g$, respectively, were obtained via an integration of $E3$.

Initial values of the quantity $d$ were calculated with the use of the equation $E2$, and its boundary values were obtained using $E4$. The~spherical shell shape of the matter distribution justifies imposing the following boundary values: $a\left(u,0\right)=1$, $Q\left(u,0\right)=\beta\left(u,0\right)=0$ and $p_1\left(u,0\right)=p_2\left(u,0\right)=x\left(u,0\right)=0$. Initial profiles of these functions were obtained using the equations $P2$, $M2$, $M1$, $S_{_{\left(Re\right)}}$, $S_{_{\left(Im\right)}}$ and $D$, respectively.

\subsection*{Employed schemes}

The numerical code was written in Fortran from scratch. Integration along the $u$-coordinate involved the 2$^{nd}$ order accurate Runge--Kutta method. Integration of the partial differential equations along advanced time was performed with the 2$^{nd}$ order accurate Adams--Bashforth--Moulton method, except the first point, where the trapezoidal rule was used.

The double null coordinates selected for the analysis ensure regular behaviour of all the evolving quantities within the computational domain. Numerical difficulties arise as the event horizon is approached, as the function $f$ diverges there. For this reason, a~relatively dense numerical grid is indispensable to determine the horizon location and to examine the behaviour of fields beyond it, especially for large $v$-coordinate values. The~efficiency of calculations was ensured by the use of an adaptive grid and performing integration with a~smaller step in problematic regions. For the gravitational collapse investigations, a~sufficient refinement algorithm is the one making the grid denser solely in the direction of retarded time~\cite{BorkowskaRogatkoModerski2011-084007}. The~determination of the area of the computational grid, where it should be denser was made through a~local error indicator. The~quantity should be bounded with the evolving quantities and change its value significantly in adequate regions. The~quantity $\frac{\Delta r}{r}$ along the $u$-coordinate meets the requirements and indicates the numerically problematic surrounding of the event horizon in spacetime~\cite{OrenPiran2003-044013}.

\subsection*{Tests of the code}

Analytical solutions do not exist for the investigated process and hence the accuracy of the code has to be checked via numerical possibilities in this regard. The~convergence tests were performed for two evolutions characterized by the following parameters. Evolution 1 was initiated with $\alpha=-\sqrt{3}$, $\gamma=-0.01$, while Evolution 2 with $\alpha=-1$, $\gamma=-1$, and in both cases $\as=\ah=0.04$. The~corresponding spacetime structures are presented in figures~\ref{fig:str-1b} and~\ref{fig:str-2b}, respectively.

To monitor the numerical outcomes convergence, the computations for Evolutions 1 and 2 were carried out on four grids with integration steps equal to multiples of $\delta=10^{-4}$. An integration step of a~particular grid was twice the size of a~denser one. The~convergence was examined on a~hypersurface of constant retarded time selected arbitrarily with $u=1$. The~chosen hypersurface was located close to the forming event horizon in the region where the adaptive mesh on neither of the grids was active, what enabled a~proper comparison of the results.

\begin{figure}[tbp]
\subfigure[][]{\includegraphics[width=0.475\textwidth]{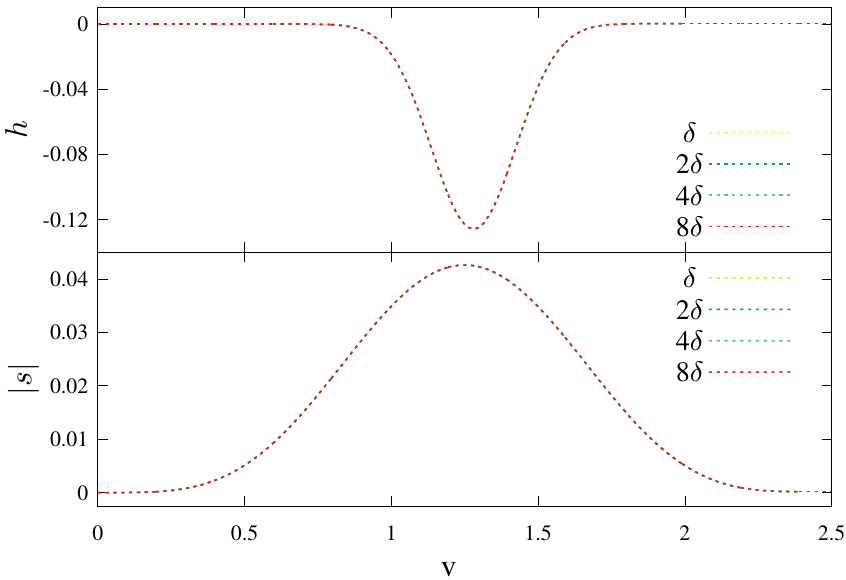}}
\hfill
\subfigure[][]{\includegraphics[width=0.475\textwidth]{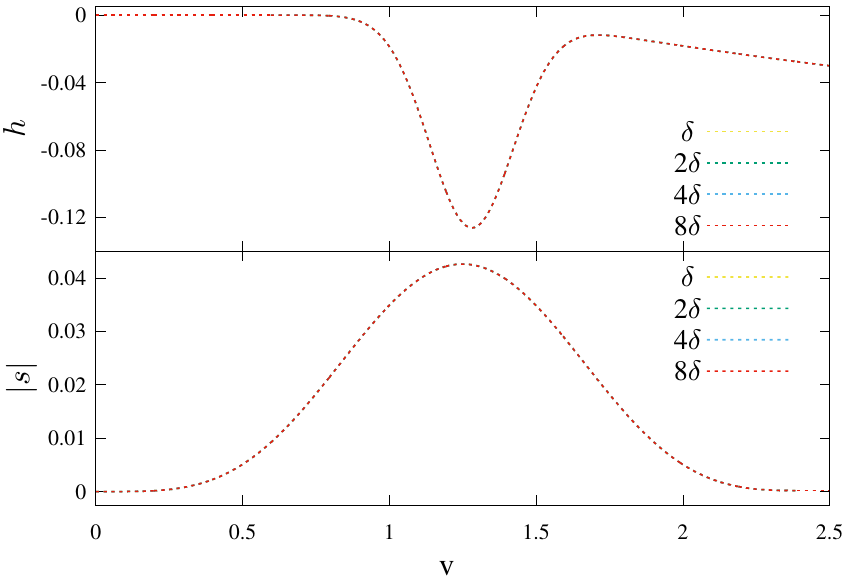}}
\caption{(color online) The~convergence of field functions. The~scalar field, $h$, and the moduli of complex scalar field, $|s|$, were plotted versus $v$ for evolutions conducted with integration steps, which were multiples of \mbox{$\delta=10^{-4}$}, along hypersurfaces of constant $u$ equal to $1$ for (a)~Evolution 1 and (b)~Evolution 2.}
\label{fig:Conv1}
\end{figure}

\begin{figure}[tbp]
\subfigure[][]{\includegraphics[width=0.475\textwidth]{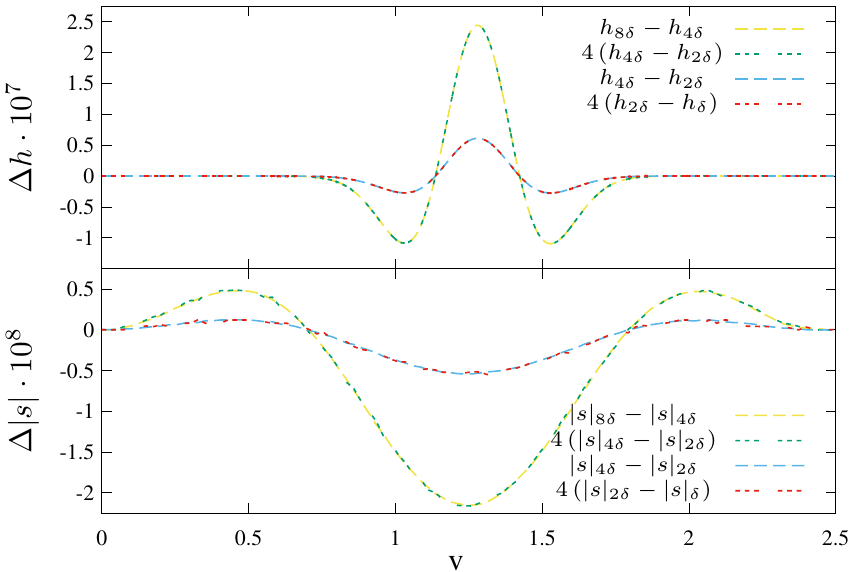}}
\hfill
\subfigure[][]{\includegraphics[width=0.475\textwidth]{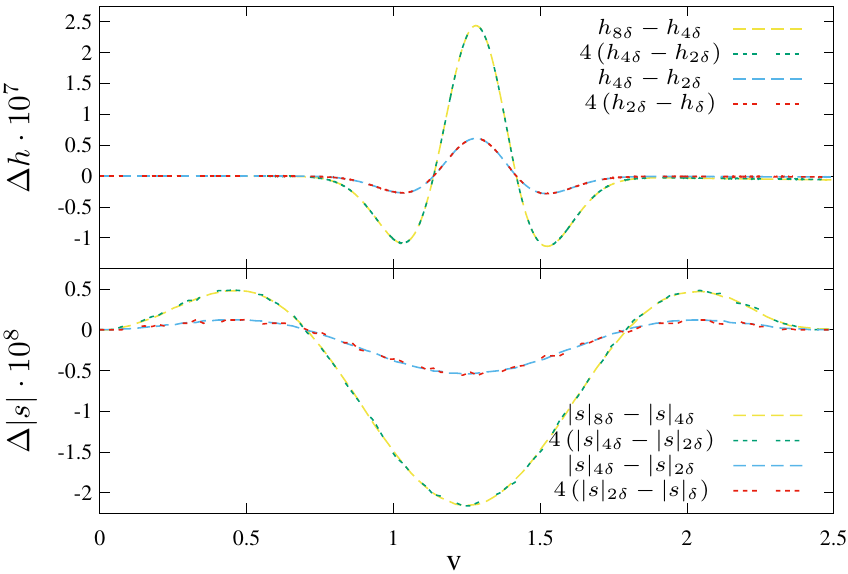}}
\caption{(color online) The~convergence of the code. The~differences between the scalar field functions, $\Delta h$, and the moduli of complex scalar field, $\Delta|s|$, calculated on grids with different integration steps (multiples of \mbox{$\delta=10^{-4}$}) and their multiples were obtained along the same hypersurfaces of constant $u$ as in figure~\ref{fig:Conv1} for (a)~Evolution 1 and (b)~Evolution 2.}
\label{fig:Conv2}
\end{figure}

The evolving field functions along the chosen hypersurface of $u=const$ from within the range of advanced time in which the functions are initially non-vanishing for all the examined integration steps are shown in figure~\ref{fig:Conv1}. The~maximal observed discrepancy between the functions calculated on the grids with the smallest and biggest steps was equal to $2.55\cdot 10^{-4}\%$. Figure~\ref{fig:Conv2} confirms the 2$^{nd}$ order convergence of the numerical code. The~maximal divergence between the field profiles obtained on two grids with a~quotient of integration steps equal to $2$ and their respective quadruples was $4\cdot 10^{-1}\%$. The~errors decreased as the grid density increased. The~overall analysis revealed that the expected convergence was achieved and both the algorithm and the numerical code were appropriate for solving the obtained system of equations~\eqref{eqn:P1-2}--\eqref{eqn:D} describing the dynamics of interest.

\acknowledgments

AN was supported by the National Science Centre, Poland, under a~postdoctoral scholarship DEC-2016/20/S/ST2/00368. 
LN was supported by the National Science Centre, Poland, under a~grant DEC-2017/26/D/ST2/00193.





\bibliographystyle{JHEP}
\bibliography{dilatonicLoveGBcollapseJHEP.bib}

\end{document}